\documentclass[pra,superscriptaddress,reprint,amsmath,amssymb,nofootinbib]{revtex4-2}

\usepackage{graphicx}
\usepackage{dcolumn}
\usepackage{subfiles} 
\usepackage{bm}
\usepackage[usenames,dvipsnames]{xcolor}
\usepackage[pdftex,plainpages=false,colorlinks=true]{hyperref}
\usepackage[toc,page,title,titletoc,header]{appendix} 
\usepackage{multirow}
\usepackage{tabularx}
\usepackage{algorithm}
\usepackage{algpseudocode}
\algnewcommand{\To}{\textbf{To }}
\algnewcommand\Input{\item[\textbf{Initialisation:}]}%
\algnewcommand\Output{\item[\textbf{Output:}]}%
\usepackage{booktabs}

\usepackage[caption=false]{subfig}

\usepackage[normalem]{ulem}

\DeclareMathOperator*{\argminA}{arg\,min}

\newcommand{\CSOneName}{\texttt{H1:PEM-CS\_MAINSMON\_EBAY\_1\_DQ}}

\newcommand{\StrainChanName}{\texttt{L1:DCS-CALIB\_STRAIN\_C01\_AR}}
\newcommand{\PEMChanName}{\texttt{L1:PEM-CS\_MAINSMON\_EBAY\_1\_DQ}}
\begin{document}
%\preprint{}

\title[test short title]{Adaptive cancellation of mains power interference in continuous gravitational wave searches with a hidden Markov model}% Force line breaks with \\
%\thanks{this would be a footnoted on the first page}%

%\input{authors.tex}

%\author{T. Kimpson$^{1,2}$, S. Suvorova, Liu, Melatos, Middleton, Evans, Moran, Meyers, any others?}

\author{T. Kimpson}
\email[]{tom.kimpson@unimelb.edu.au}
\affiliation{School of Physics, University of Melbourne, Parkville, VIC 3010, Australia} %
\affiliation{OzGrav, University of Melbourne, Parkville, VIC 3010, Australia}

\author{S. Suvorova}
\affiliation{Department of Electrical and Electronic Engineering, University of Melbourne, Parkville, Victoria 3010, Australia}

\author{H.Middleton}
\affiliation{School of Physics and Astronomy, University of Birmingham, Edgbaston, Birmingham, B15 2TT, United Kingdom}
\affiliation{Institute for Gravitational Wave Astronomy, University of Birmingham, Edgbaston, Birmingham, B15 2TT, United Kingdom}

\author{C.Liu}
\affiliation{OzGrav, University of Melbourne, Parkville, VIC 3010, Australia}
\affiliation{Department of Electrical and Electronic Engineering, University of Melbourne, Parkville, Victoria 3010, Australia}

\author{A. Melatos}
\affiliation{School of Physics, University of Melbourne, Parkville, VIC 3010, Australia} %
\affiliation{OzGrav, University of Melbourne, Parkville, VIC 3010, Australia}

\author{R. Evans}
\affiliation{OzGrav, University of Melbourne, Parkville, VIC 3010, Australia}
\affiliation{Department of Electrical and Electronic Engineering, University of Melbourne, Parkville, Victoria 3010, Australia}

\author{W. Moran}
\affiliation{Department of Electrical and Electronic Engineering, University of Melbourne, Parkville, Victoria 3010, Australia}

%\affiliation{School of Physics, University of Melbourne, Parkville, VIC 3010, Australia} %
%\affiliation{OzGrav, University of Melbourne, Parkville, VIC 3010, Australia}
%\affiliation{Department of Electrical and Electronic Engineering, University of Melbourne, Parkville, Victoria 3010, Australia}

%\author[Kimpson]{Tom Kimpson$^{1,2}$\thanks{Contact e-mail: \href{tom.kimpson@unimelb.edu.au}{tom.kimpson@unimelb.edu.au}}, Andrew Melatos$^{1,2}$, Joseph O'Leary$^{1,2}$, Julian B. Carlin$^{1,2}$, Robin J. Evans$^{3}$, \newauthor William Moran$^{3}$, Tong Cheunchitra$^{1,2}$, Wenhao Dong$^{1,2}$, Liam Dunn$^{1,2}$, Julian Greentree$^{3}$, Nicholas J. O'Neill$^{1,2}$, \newauthor Sofia Suvorova$^{3}$, Kok Hong Thong$^{1,2}$, Andrés F. Vargas$^{1,2}$%
%	%\thanks{Present address: Science magazine, AAAS Science International, \mbox{82-88}~Hills Road, Cambridge CB2~1LQ, UK}%
%	\\
%	% List of institutions
%	$^{1}$School of Physics, University of Melbourne, Parkville, VIC 3010, Australia \\
%	$^{2}$OzGrav, University of Melbourne, Parkville, VIC 3010, Australia \\
%	$^{3}$Department of Electrical and Electronic Engineering, University of Melbourne, Parkville, Victoria 3010, Australia }

\date{\today}% It is always \today, today,

\begin{abstract}
		Continuous gravitational wave searches with terrestrial, long-baseline interferometers are hampered by long-lived, narrowband features in the power spectral density of the detector noise, known as lines. Candidate GW signals which overlap spectrally with known lines are typically vetoed. Here we demonstrate a line subtraction method based on adaptive noise cancellation, using a recursive least squares algorithm, a common approach in electrical engineering applications such as audio and biomedical signal processing. We validate the line subtraction method by combining it with a hidden Markov model (HMM), a standard continuous wave search tool, to detect an injected continuous wave signal with an unknown and randomly wandering frequency, which overlaps with the mains power line at $60 \, {\rm Hz}$ in the Laser Interferometer Gravitational Wave Observatory (LIGO). The performance of the line subtraction method is tested on a injected continuous wave signal obscured by (a) synthetic noise data with both Gaussian and non-Gaussian components, and (b) real noise data obtained from the LIGO Livingston detector. In both cases, before applying the line subtraction method the HMM does not detect the injected continuous wave signal. After applying the line subtraction method the mains power line is suppressed by 20--40 dB, and the HMM detects the underlying signal, with a time-averaged root-mean-square error in the frequency estimate of $\sim 0.05 $ Hz. The performance of the line subtraction method with respect to the characteristics of the 60 Hz line and the control parameters of the recursive least squares algorithm is quantified in terms of receiver operating characteristic curves.

\begin{description}
\item[DOI]
\end{description}
\end{abstract}

\maketitle

\section{Introduction} \label{sec:intro}
Instrumental noise artifacts in gravitational wave (GW) searches with terrestrial, long-baseline interferometers \cite{Aasi2015,AcerneseEtAlAdVirgo:2015,Akutsu2021PTEP} are classified according to their duration and spectral properties.  Short-lived, non-stationary, recurrent noise events such as optomechanical glitches typically last for seconds and exhibit distinctive spectral signatures, e.g they can be chirp-like \cite{DetCharGW150914:2016,Davis2021,Davisgalaxies10010012,Glanzer2023}. Long-lived, quasi-stationary, broadband noise sources include seismic disturbances at low frequencies, test mass thermal fluctuations at intermediate frequencies, and photon shot noise at high frequencies \cite{AasiEtAlAdLIGO:2015,LIGOnoiseguide,2020Galax...8...82F,Nguyen2021}. Long-lived narrowband spectral artifacts --- termed instrumental lines --- are caused by electrical subsystems (e.g. mains power, clocks, oscillators), mechanical subsystems (e.g. test mass and beam-splitter violin modes) and calibration processes \cite{CovasEtAl:2018}, although sometimes the origin of a specific feature is unknown. Instrumental lines are disruptive, especially for continuous wave (CW) searches where the target astrophysical signal is quasi-monochromatic and resembles the noise artifact spectrally. Many above-threshold candidates discovered in published CW searches are vetoed because they coincide with known instrumental lines \cite{Piccinni2022,Riles2023,Wette2023}, for instance, in data from Observing Run 3 (O3) with the Laser Interferometer Gravitational Wave Observatory (LIGO), Virgo and the Kamioka Gravitational Wave Detector (KAGRA)  \cite[e.g.][]{Ligo_lineveto1,ligo_lineveto2,ligo_lineveto3} \newline

Several techniques have been implemented by the LIGO-Virgo-KAGRA (LVK) Collaboration to identify, characterize and suppress instrumental noise artifacts \cite{Davis2021,Davisgalaxies10010012}. Some techniques identify and veto an artifact (or gate data segments) based on its time-frequency signature \cite{Abbott2018CQVeto,Steltner2022Ph}. Other techniques perform offline noise subtraction with reference to auxiliary data from physical environmental monitors (PEMs) \cite{Driggers2012RScI,Tiwari2015,Davis2019,Driggers2019}. PEMs can be used to witness correlated noise and generate a reference signal directly or elucidate and quantify multichannel couplings \cite{Jung2022PRD,marin1997}. Additionally some techniques are based on machine learning \cite{Cuoco2020,Vajente2020PhR,OrmistonEtAl:2020,Yamamoto2023} or multi-detector consistency \cite{KeitelEtAl:2014, 2015PhyS...90l5001L}. In CW searches specifically, the distinctive amplitude and frequency (Doppler) modulations associated with the Earth's rotation and revolution respectively can be exploited to discriminate between terrestrial noise artifacts and astrophysical signals \cite{Zhu2017PhRvD,JonesPhysRevD.106.123011}.  \newline

In most  of the situations above, the practical effect of an instrumental line is to excise the relevant part of the observing band from a CW search. That is, if an above-threshold CW search candidate coincides with a known instrumental line, the candidate is often vetoed under current practice without further analysis, e.g.\ without comparing the expected strength of the noise line with the measured strength of the candidate \footnote{A regularly updated log of narrowband instrumental lines in the LVK detector is maintained at \href{https://dcc.ligo.org/LIGO-T2100200/public}{dcc.ligo.org/LIGO-T2100200/public} for public reference.}, although some line robust algorithms exist which combine statistics from different detectors to suppress single detector line artifacts \cite{KeitelEtAl:2014,2015CQGra..32c5004K,PhysRevD.93.084024,Keitel_2016}; we refer the reader to \cite{2022PhRvD.106j2008A} for an overview of different continuous wave search pipelines and the associated vetoing procedures. In this paper we take a step towards lifting the above limitation. We introduce an adaptive noise cancellation (ANC) scheme  based on an adaptive recursive least squares (ARLS) algorithm which suppresses narrowband noise proportional to a known PEM reference signal. We then apply the ANC scheme to a CW search algorithm based on a hidden Markov model (HMM), which detects and tracks quasi-monochromatic GW signals with wandering frequency and has been tested and validated thoroughly in multiple LVK searches \cite{Suvorova2016PhRv,Piccinni2022,Riles2023,Wette2023}. We demonstrate with synthetic data that the ANC scheme and HMM algorithm together can successfully detect a GW signal overlapping the 60 Hz mains power line, if the signal exceeds a well-defined minimum amplitude. The approach extends naturally to other instrumental lines, provided they have an associated witness PEM channel, a topic for future work. \newline 

The paper is organized as follows. In Section \ref{sec:pgi} we outline a mathematical model for the 60 Hz mains power line and the PEM reference signal, and justify the assumptions of the model by reference to data from the LIGO Livingston interferometer.  In Section \ref{sec:method} we introduce ANC formulated as an ARLS method. In Section \ref{sec:results} we validate the ANC scheme on synthetic GW strain data, in conjunction with a pre-existing HMM search pipeline, and demonstrate the successful recovery of a frequency-wandering CW signal. In Section \ref{sec:theroccurves} we quantify the performance of the HMM pipeline and ANC scheme as a function of the key parameters of the mains power interference and ANC filter. The validation tests are extended to real LIGO noise in Section \ref{sec:real_data}. Concluding remarks are made in Section \ref{sec:conclusions}.

\section{Mains power interference} \label{sec:pgi}
The goal of this paper is to detect a quasi-monochromatic GW signal in a data stream contaminated by two kinds of noise: additive Gaussian noise which is fundamental and irreducible, and additive non-Gaussian interference from a long-lived narrow spectral feature which can be filtered out in principle given an accurately measured reference signal. For this work we consider the spectral line at 60 Hz that results from the mains electricity in North America as the additive non-Gaussian interference, as it is a long-standing impediment to many CW searches with the LIGO interferometers. Similar lines due to mains electricity arise at 50 Hz for Virgo \citep{Virgo2012_lines} and 60 Hz for KAGRA \citep{Kagra2023_lines}. In Section \ref{sec22} we briefly review the 60 Hz LIGO interference line before proceeding in Section \ref{sec21} to specify the assumed mathematical forms of the interference and reference signals. In Section \ref{sec23} we justify the assumptions of Section \ref{sec21} by analysing differential arm length (DARM channel) and environmental (mains power monitoring) data from the LIGO interferometers.

\subsection{LIGO 60 Hz Interference}  \label{sec22}
%Useful literature: https://journals.aps.org/prd/pdf/10.1103/PhysRevD.97.082002

\begin{figure}
	\begin{center}
		\includegraphics[width=\columnwidth]{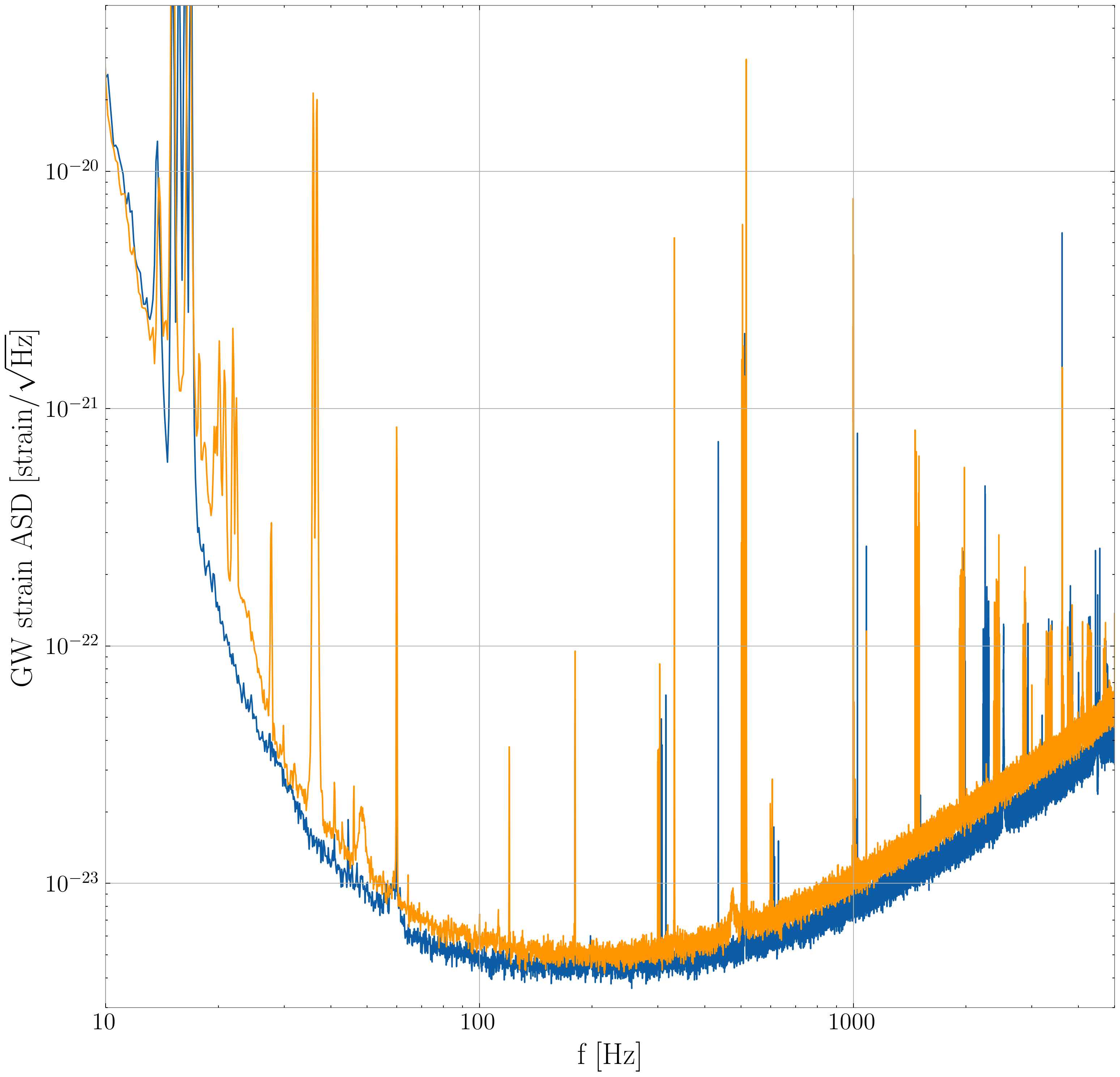}
	\end{center}
	\caption{GW strain amplitude spectral density (abbreviated as ASD on axis label) for LIGO-Hanford (orange) and LIGO-Livingston (blue) for a snapshot of data ($10.0$ minutes) from O3 (channel \texttt{*:DCS-CALIB\_STRAIN\_C01\_AR}, see Ref.~\cite{LIGO_O3, GWOSC:online}. The vertical axis plots the amplitude spectral density of the detector noise in units of Hz$^{-1/2}$, while the horizontal axis plots the observation frequency in units of Hz. The spectral line at $60\,{\rm Hz}$ is clearly visible as a spike rising three decades, along with multiple other instrumental lines at other frequencies.}\label{fig:strainSensitivity}
\end{figure}

LIGO data contains multiple long-duration narrow lines in addition to Gaussian noise. These lines are shown in Figure \ref{fig:strainSensitivity} where we plot the amplitude spectral density of the GW strain channel \texttt{*:DCS-CALIB\_STRAIN\_C01\_AR} for both LIGO-Hanford (\texttt{H1}, orange curve) and LIGO-Livingston (\texttt{L1}, blue curve). The small amplitude fluctuations are the result of Gaussian noise whereas the large amplitude spikes are the long-duration lines. The provision of mains power electricity in North America via an alternating current (AC) with frequency 60 Hz leads to a distinct line in Figure  \ref{fig:strainSensitivity}. The mains power couples to the GW strain channel, because the performance of the high-sensitivity electronic components within LIGO varies with respect to the input power voltage. Additionally, the magnetic fields that arise from the AC mains supply couple to magnets on optical components \citep{CovasEtAl:2018}. \newline 

\begin{figure}
	\includegraphics[width=\columnwidth]{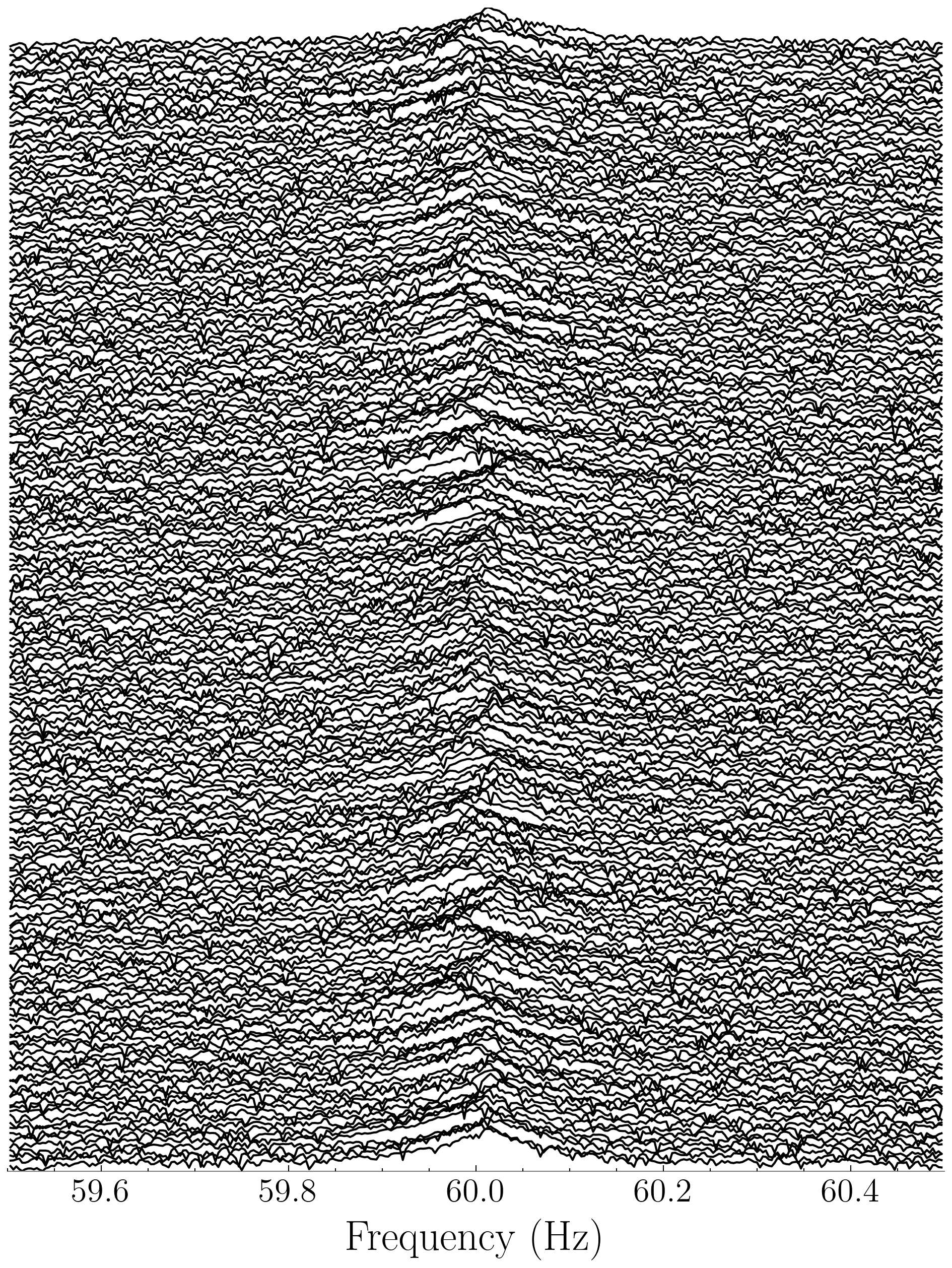}
	\caption{Cascade plot showing the temporal evolution of the amplitude spectral density (units: Hz$^{-1/2}$) of the LIGO-Hanford three-phase mains-power PEM monitor \CSOneName~ (corner station, phase 1) as a function of frequency (units: Hz). Each trace corresponds to $320\,{\rm s}$ ($\approx 5\,{\rm min}$) of data ($210$ traces stacked vertically). The peak of the mains power line wanders by $\approx \pm 30$ mHz about its time-averaged frequency $60.004$ Hz.}
	\label{fig:powerCascade}
\end{figure}

Some spectral lines are static, but the 60 Hz line wanders in time, due to variations in the load in the North American power grid. The wandering is displayed in the cascade plot in Figure \ref{fig:powerCascade}, where the amplitude spectral density of the output signal of the LIGO-Hanford PEM channel \CSOneName \, is plotted for 210 5-min samples stacked vertically. The peak of the amplitude spectral density varies from 59.969 Hz  to 60.053 Hz, across the 210 samples with an average value of 60.004 Hz. Its full width half maximum varies from 5.294 mHz to 29.377 mHz, with an average value of 10.749  mHz. For a full review of LIGO spectral artifacts, including the 60 Hz line, we refer the reader to \cite{CovasEtAl:2018}. \newline 

The $60 \, {\rm Hz}$ line occurs in the neighbourhood of plausible astrophysical signals. For example, mass quadrupole radiation from a rotating mountain on the Crab pulsar would be emitted at twice the star's spin frequency, viz. $59.89 \, {\rm Hz}$ (epoch 48442.5 MJD) \cite{crab,psrqpy}.

%https://arxiv.org/abs/1903.03866 useful for Jaranowski discussion
\subsection{Statement of the problem: signal and noise models}  \label{sec21}
Let $x(t)$ denote the scalar time series output by the GW strain channel of a LIGO-like long baseline interferometer. Suppose that $x(t)$ is sampled at discrete times $t_n$, with $1 \leq n \leq N$ and uniform sampling interval $\Delta t = t_n - t_{n-1}$. Let $r(t)$ denote the scalar time series output by the environmental reference channel relevant for filtering interference; here $r(t)$ is of one of the three phases of mains power measured at some reference point in the detector. The reference signal is usually sampled less frequently than $x(t)$ at discrete times $t_{n_k}$, with $1 \leq k \leq K$ and $1 \leq n_k \leq N$. We assume for the sake of convenience that every $t_{n_k}$ coincides with some $t_n$ for all $k$, but the condition is not essential. \newline 

The strain channel is composed of a GW signal $h(t)$, non-Gaussian interference $c(t)$ (sometimes called ``clutter") and Gaussian noise $n(t)$ in a linear combination,
\begin{eqnarray}
	x(t) = h(t) + c(t) + n(t) \ .
	\label{eq:data}
\end{eqnarray}
 In this paper the GW signal takes the form predicted by \citet{Jaranowski1998} for a biaxial rotor, e.g. a neutron star (NS) emitting gravitational waves at multiples of the star's spin frequency $f_{\star}$. The GW signal is quasi-monochromatic, and modulated in both amplitude and frequency by the rotation of the Earth and the Earth's orbital motion. The noise $n(t)$ is white with $\langle n(t_n) n(t_{n'})\rangle = \sigma_n^2 \delta_{n n'}$. Noise samples $n(t_n)$ are drawn from a Gaussian distribution with zero mean and variance $\sigma_n^2$. As one can see in Figure \ref{fig:strainSensitivity}, the noise is generally not white. However, whitening procedures can easily be used in pre-processing to make the noise approximately agree with this assumption. The interference $c(t)$ takes a form determined by instrumental processes but is generally a long-lived, narrowband spectral feature. We relate $c(t)$ to the mains voltage $r(t)$ in this paper. \newline

 The statistical properties of the mains power grid are complicated, and there is no predictive, authoritative, first principles model to describe them \cite{888ed69a34a94bd1bdcb5073edcdd3fa,8571793}. In particular, the statistics of frequency fluctuations reflect hard-to-predict fluctuations in demand, which stem from random exogenous spatio-temporal factors, such as weather, economic activity, faults, and so on. \footnote{The power grid responds traditionally to demand by adjusting the frequency. When load increases, the frequency drops, more fuel is supplied to turbine-based generators to meet the increased load, and the frequency rises.} We therefore rely on a tunable, phenomenological model to capture three important qualitative features of mains power. First, the frequency is maintained at a constant value across the grid to a good approximation by internal grid mechanisms (effectively a phase locked loop), with central frequency $f_{\rm ac} = 60$ Hz in North America. A slow periodic modulation occurs around $f_{\rm ac}$ with a small amplitude $\Delta f_{\rm ac} \lesssim 0.5$ Hz and period $P$, which wanders randomly and uniformly in the range $0 \leq P \leq P_{\rm max}$. Second, the phase offset $\Theta(t)$ of the voltage $r(t)$ wanders stochastically. We assume that the phase noise is white and Gaussian, with $n_{\Theta}(t_n)$ drawn from a Gaussian distribution with zero mean and variance $\sigma_{\Theta}^2$, i.e. $\langle n_{\Theta} (t_n) n_{\Theta} (t_{n'})\rangle = \sigma^2_{\Theta} \delta_{n n'}$. Third, the voltage amplitude, $A_r(t)$, is random. We assume that samples $A_r(t_n)$ are distributed uniformly within $[A_{\rm ac} - \Delta A_{\rm ac} , A_{\rm ac} + \Delta A_{\rm ac}]$. We can then write the reference voltage as,
 \begin{eqnarray}
 	r(t) = A_r(t_n) \cos \left[ 2 \pi f_{\rm ac} t + \Theta(t)\right] +n_r(t_n) \ ,
 	\label{eq:voltage}
 \end{eqnarray}
with
 \begin{eqnarray}
\Theta(t) = 2 \pi \Delta f_{\rm ac} \cos\left[\frac{2 \pi t}{P(t_n)}\right] + n_{\Theta} (t_n) \ ,
\label{eq:voltage_theta}
\end{eqnarray}
%\begin{align}
%	r(t) &= \nonumber \\
%	& A_r(t_n) \cos \left( 2 \pi f_{\rm ac} t + 2 \pi \Delta f_{\rm ac} \cos\left(\frac{2 \pi t}{P(t_n)}\right) + n_{\Theta} (t_n)\right) \nonumber \\
%	& +w(t_n)
%	\label{eq:voltage}
%\end{align}
for $t_n \leq t \leq t_{n+1}$. That is, at time $t_n$, random variables $A_r(t_n)$, $P(t_n)$ and $n_{\Theta} (t_n)$ are drawn from the distributions $\mathcal{U}[A_{\rm ac} - \Delta A_{\rm ac},A_{\rm ac} + \Delta A_{\rm ac} ]$, $\mathcal{U}[0, P_{\rm max}]$, and $\mathcal{N} [0, \sigma_{\Theta}^2]$ respectively, where $\mathcal{U}$ denotes uniform, and $\mathcal{N}$ denotes normal. Equation \eqref{eq:voltage} then steps forward over a time interval $\Delta t$; that is, $r(t)$ is discontinuous from one time step to the next due to random sampling. In Equation \eqref{eq:voltage}, $n_r(t_n)$ is the reference signal measurement noise at $t_n$, assumed to be white and Gaussian with $n_r(t_n)$ drawn from a Gaussian with zero mean and variance $\sigma_r^2$. All the white measurement and process noises are assumed to be independent for simplicity. \newline 

Mains power couples into the strain channel in various complicated ways, e.g. through electronic devices, or inductively through ambient magnetic fields. A central assumption in this work is that the interference in the strain channel is an exact, amplitude-scaled replica of the reference signal up to a delay $\tau_{\rm delay}$ which is attributed to spatial propagation between the PEM position and the GW-sensing apparatus. Hence we can express the interference as 
 \begin{align}
	c(t) = A_c(t_n - \tau_{\rm delay}) \cos \biggl [ & 2 \pi f_{\rm ac} (t_n - \tau_{\rm delay}) \nonumber \\ 
	&+ \Theta(t_n - \tau_{\rm delay}	) \biggr ] \ ,
	\label{eq:vclutter}
\end{align}
for $t_n \leq t \leq t_{n+1}$. The amplitude $A_c$ is distributed as $\mathcal{U}[A'_{\rm ac}- \Delta A'_{\rm ac} ,A'_{\rm ac} + \Delta A'_{\rm ac} ]$. For this work we consider $0 \leq \tau_{\rm delay} \leq 100 \Delta t$, but wider or narrower ranges are possible and straightforward to implement. The assumption that $c(t)$ is a scaled replica of $r(t)$ is tested in Section \ref{sec23}. Alternative coupling mechanisms which are recorded by PEMs in different positions may have different $\tau$ values, potentially leading to stochastic overlap of different time-delay components.

\subsection{Coherence between $x(t)$ and $r(t)$}  \label{sec23}
A key assumption in Section \ref{sec21} is that the 60 Hz interference recorded in the reference PEM channel is also present in the strain channel. That is, the mains voltage recorded by the PEM is imprinted onto the strain channel up to a proportionality constant and time delay. In order to test this assumption we calculate the coherence between the strain channel and the PEM channel. \newline 

The coherence between two time series $x(t)$ and $r(t)$ is given by 
\begin{equation}
	C_{xr}(f)	= \frac{|P_{xr}(f)|^2}{P_{xx}(f) P_{rr}(f)} \ , \label{eq:coherence}
\end{equation}
where $f$ is the Fourier frequency, $P_{xr}$ is the (cross) power spectral density
\begin{equation}
	P_{xr}(f)	= \int_{-\infty}^{\infty} d \tau R_{xr}(\tau) e^{-2 \pi i f \tau }  \ ,    \label{eq:psd}
\end{equation}
in terms of the cross-correlation function,
\begin{equation}
	R_{xr}(\tau) = \int_{-\infty}^{\infty} dt x(t) r(t+ \tau)  \ ,  \label{eq:cross-corr}
\end{equation}
and $P_{xx}(f)$, $P_{rr}(f)$, $R_{xx}(f)$ and $R_{rr}(f)$ are defined analogously to Equations \eqref{eq:psd} and \eqref{eq:cross-corr}. In Equation \eqref{eq:coherence} we have $0 \leq C_{xr}(f) \leq 1$. $C_{xr}(f) = 0$ indicates that the two signals are completely unrelated, whilst $C_{xr}(f) = 1$ indicates that the two signals have an ideal, noiseless, linear relationship, i.e. the convolution $x(t) = g(t) \ast r(t)$ for impulse response function $g(t)$.  The intermediate situation $ 0 < C_{xr}(f) < 1$ indicates that the relationship between $x(t)$ and $r(t)$ is non-linear, either due to measurement noise or contributions to $x(t)$ from additional signals. Heuristically, for linear systems, $C_{xr}(f)$ can be understood as the fraction of the power in $x(t)$ that is produced by $r(t)$, at Fourier frequency $f$. \newline

We use data for the strain and PEM channels from the first part of the third LIGO observing run, O3a \cite{LIGO_O3}. These data are obtained via the LIGO Data Grid  \footnote{\url{https://computing.docs.ligo.org/guide/computing-centres/ldg/}} using the \texttt{GWPy} package \cite{gwpy}. In O3a there are nine independent PEM channels at both LIGO-Livingston and LIGO-Hanford that directly measure the mains voltage \footnote{\url{https://git.ligo.org/detchar/ligo-channel-lists/-/blob/master/O3/L1-O3-pem.ini}. See also \cite{pemrefs}.}. In this paper we consider just the LIGO-Livingston data, as they suffice to make the point. Specifically, there are 3 PEM mains voltage monitors in the electronics bay in each of the X-arm end station (\texttt{EX}), the Y-arm end station (\texttt{EY}) and the corner station (\texttt{CS}). Each of the three PEMs at each station measures a separate component of three-phase mains power. The strain data are sampled at 16384 Hz and downsampled to 1024 Hz to match the sampling rate of the PEM channels. \newline 
%https://arxiv.org/pdf/1812.05225.pdf

\begin{figure}
	\begin{center}
		\includegraphics[width=\columnwidth]{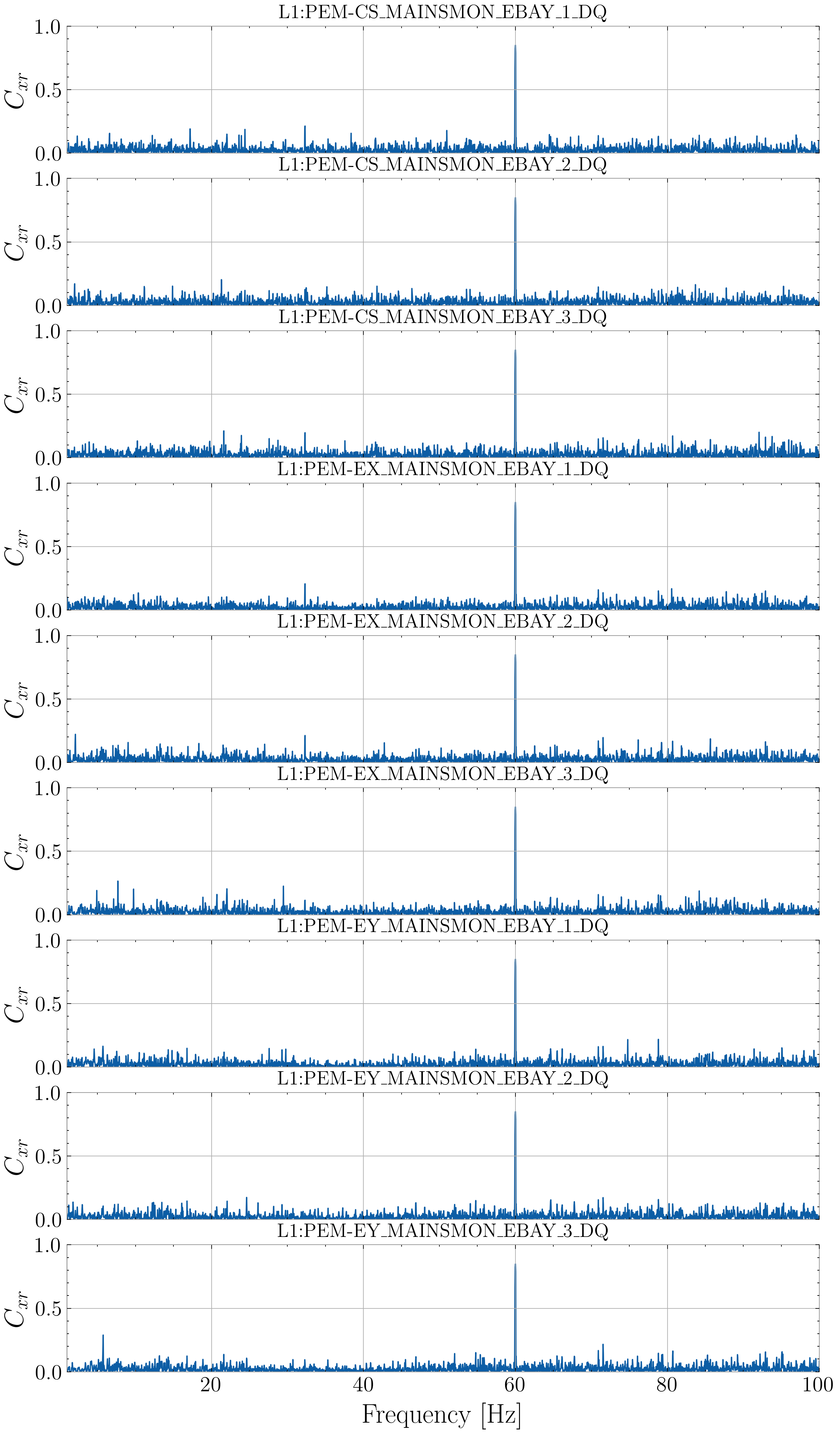}
	\end{center}
	\caption{\label{coherenceplot_1} Coherence $C_{xr}$, defined by Equation \eqref{eq:coherence}, relating the LIGO-Livingston strain channel  \texttt{L1:DCS-CALIB\_STRAIN\_C01\_AR} and the nine three-phase mains-power PEM channels at the LIGO-Livingston end stations and corner station during a 10 minute observation interval. Clear features at 60 Hz are present in every panel, with amplitudes $ > 0.8$ (unitless).}
\end{figure}

Figure \ref{coherenceplot_1} displays $C_{xr}(f)$ relating the high latency, calibrated strain channel \texttt{L1:DCS-CALIB\_STRAIN\_C01\_AR} and each of the nine reference PEM channels over a 10 minute interval. For all channels there is a clear spike in $C_{xr}(f)$ at 60 Hz, with a mean amplitude $=0.85$ (unitless). \newline 

\begin{figure*}
	\begin{center}
		\includegraphics[width=\textwidth]{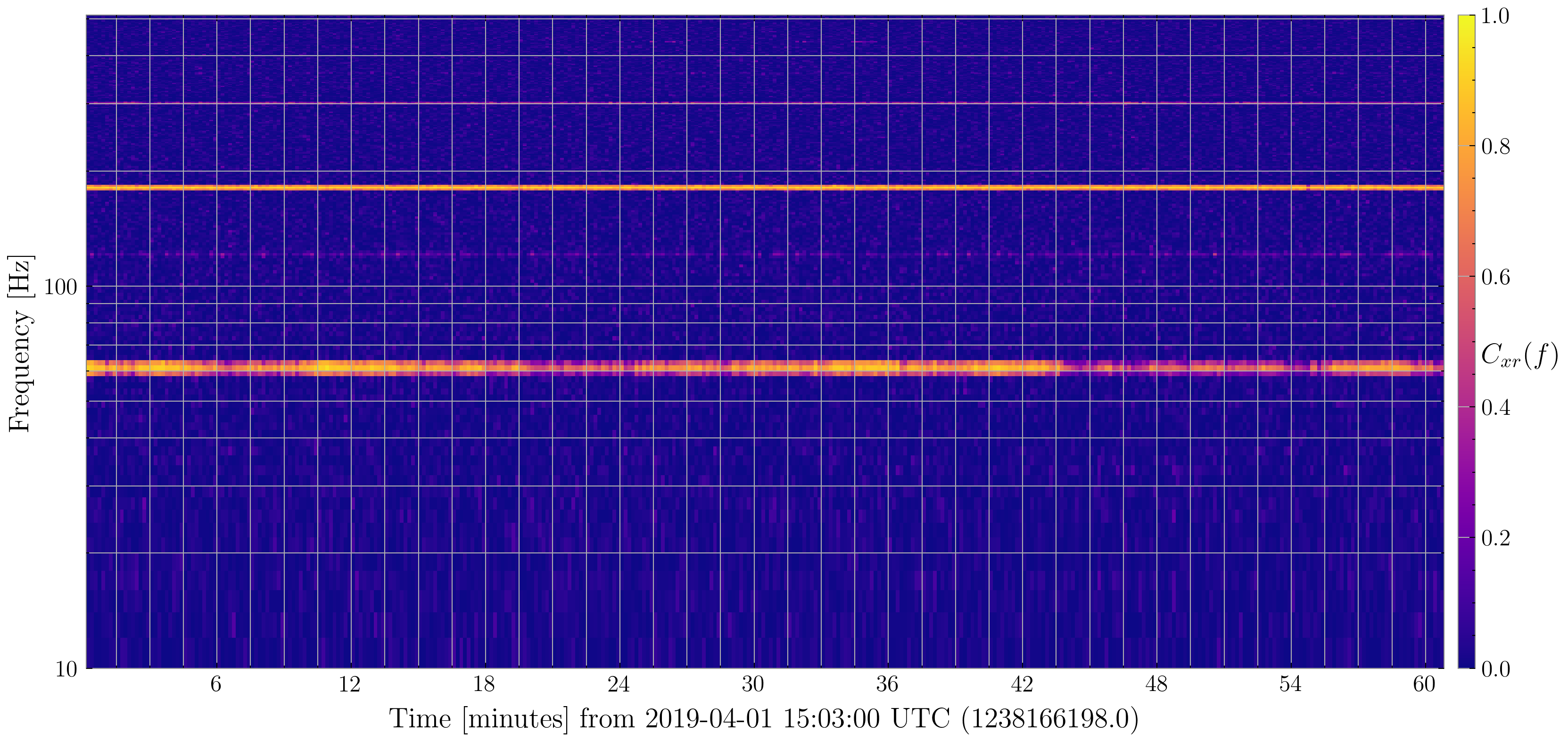}
	\end{center}
	\caption{\label{correlation_1}
		Coherence spectrogram between the strain channel \StrainChanName  \, and the PEM channel \PEMChanName  	\, covering a 1 hour period of O3 data. The horizontal axis is the time in minutes from the start of O3a, divided into 10-s bins. The vertical axis is the Fourier frequency in Hz. The pixel colour denotes the coherence $C_{xr}$, defined by Equation \eqref{eq:coherence}, in that time-frequency bin; see colour bar at right (dimensionless). High coherence is observed at 60 Hz due to the mains power interference. Additional coherence features are visible at 0.2 kHz and 0.3 kHz, corresponding to higher harmonics of the mains interference.}
\end{figure*}

 It is important to note that whilst $C_{xr}(f=60 \text{Hz})$ is strong, it also varies with time. Figure \ref{correlation_1} displays the coherence spectrogram, i.e. the time-varying coherence, between the strain channel and the PEM channel \PEMChanName \, (c.f. top panel, Figure \ref{coherenceplot_1}) over a 1-hr time interval (c.f. the 10 minute interval of Figure \ref{coherenceplot_1}). $C_{xr}(f)$ is calculated in 10-s blocks; that is, every pixel in Figure \ref{correlation_1} is 10 s wide horizontally. We use a Fourier transform window of length 0.5 s, with each window overlapping by 0.25 s (c.f. Welch's method, Ref. \citep{Welch1161901}). The results resemble Figure \ref{coherenceplot_1}; there is a spectral feature with high coherence ($C_{xr}(f=60 \rm Hz) = 0.73$ on average) which persists across the full 1-hr time interval. However, whilst the feature is persistent, $C_{xr}(f=60 \text{Hz})$ is not constant across the entire interval. Instead it changes in time, with variance $=0.018$. These variations are visible in the changing colours of the spectrogram at 60 Hz, leading to ``patchiness" of the line. Additional features in the $C_{xr}(f)$ spectrum are seen near 0.2 kHz and 0.3 kHz, corresponding to higher harmonics of the mains interference. Similar results are obtained for the other PEM channels shown in Figure \ref{coherenceplot_1}. Figures  \ref{coherenceplot_1} and  \ref{correlation_1} justify in part the assumptions in Section \ref{sec21}. \newline

What is the threshold value of $C_{xr}(f=60 \text{Hz})$ required for noise cancellation to work sufficiently, such that the HMM presented in Section \ref {sec:results} can successfully recover the underlying signal? This is an empirical question which depends on the properties of the underlying signal. The coherence is related to the degree of noise cancellation by
 \begin{equation}
	R(f) = \frac{1}{1 - |C_{xr}(f)|}
\end{equation} 
 \cite{s22176591,625628}.  We show in the validation tests on real LIGO data presented in Section \ref{sec:real_data}, that the observed level of coherence, viz. $C_{xr}(f=60 \text{Hz}) \approx 0.85$ at 60 Hz, allows the ANC scheme to suppress the mains power interference sufficiently for the HMM to detect an underlying signal, whose amplitude is comparable to detectable signals away from the 60-${\rm Hz}$ line.


\section{ANC scheme}\label{sec:method}

ANC is a method for estimating an underlying signal corrupted or obscured by additive noise, which is itself estimated independently \cite{Widrow1451965}. In contrast to other common optimal filtering methods (e.g. Wiener, Kalman) ANC assumes no \textit{a priori} model of either the signal or the noise. Instead, ANC makes use of a reference input, which is correlated in some unknown way with the noise in the primary data. The reference is filtered and subtracted from the primary data so as to recover the underlying signal. For our purposes, the primary data is the gravitational-wave strain channel, $x(t)$, described by Equation \eqref{eq:data}, and the reference is a PEM recording a mains voltage, $r(t)$, described by Equation \eqref{eq:voltage}. The objective of ANC is to remove the clutter $c(t)$ in Equation \eqref{eq:data} with the aid of $r(t)$ whilst leaving $h(t)$ intact. \newline 

\subsection{Filter}

The first step in implementing ANC is to convert $r(t)$ into an estimate of the clutter, $\hat{c}(t)$. The clutter estimate is subtracted from the primary signal in the time domain, defining a residual
\begin{eqnarray}
	e(t) = x(t) - \hat{c}(t) \ .\label{eq:error_estimate}
\end{eqnarray}
In the limit $\hat{c}(t) \to c(t)$, one obtains	$e(t) \to h(t) + n(t)$ and recovers a noise-cancelled timeseries. The clutter estimate is modelled by a finite duration impulse response (FIR) filter,
\begin{eqnarray}
	\hat{c}_k = \mathbf{w}^{\rm T}\mathbf{u}_k \ . \label{clutter_estimate}
\end{eqnarray}
In Equation \eqref{clutter_estimate}, ${\hat c}_k = {\hat c}(t_{n_k})$ is the clutter estimate at the $n_k$-th time step , $\mathbf{u}_k$ is the tap-input vector composed of $M$ running samples of the reference signal arranged backwards in time,
 \begin{eqnarray}
 	\mathbf{u}_k &= [r_k, r_{k-1}, \dots, r_{k-M+1}] \ ,
 \end{eqnarray}
and $\mathbf{w}$ is the tap-weight vector,
 \begin{eqnarray}
	\mathbf{w} = [w_1, w_{2}, \dots, w_{M}] \ .
\end{eqnarray}

The goal of ANC is to determine the optimal tap weights $\mathbf{w}_{\rm opt}$, that minimise the mean square error cost function,
\begin{eqnarray}
\mathbf{w}_{\rm opt} =\argminA_{\mathbf{w}}  \sum_{k=1}^{K} \lambda^{K-k-1} |e_k \left(\mathbf{w}\right)|^2 \ , \label{eq:minim}
\end{eqnarray}
where $0 \leq \lambda \leq 1$ is a ``forgetting factor" which gives exponentially less weight to older samples and can be freely chosen. The choice of $\lambda$ in this paper is discussed in Section \ref{sec:ARLS}. The minimization problem described by Equation \eqref{eq:minim} can be solved in many ways. Here we implement an ARLS algorithm. The algorithm is outlined in Section \ref{sec:ARLS}.

\subsection{ARLS algorithm}
\label{sec:ARLS}

As its name implies, the ARLS algorithm is adaptive, in the sense that the weights $\mathbf{w}$ are continually updated based on new data. It is also recursive, in the sense that $\mathbf{w}$ is updated iteratively as new data arrive, rather than reprocessing all previous data. Abbreviated pseudocode is presented below for the sake of reproducibility and the reader's convenience. The reader is referred to standard signal processing textbooks, e.g. \cite{HaykinAdaptiveFT:2002} for more information. \newline 

\begin{figure}
	\begin{center}
		\includegraphics[width=\columnwidth]{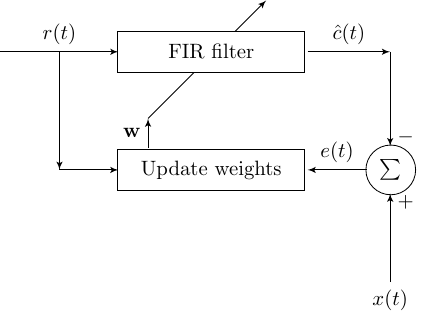}
	\end{center}
	\caption{Block diagram of the ARLS method described in Section \ref{sec:ARLS}. The reference signal $r(t)$ is passed to an FIR filter to construct an estimate $\hat{c}(t)$ of the clutter. Subtracting $\hat{c}(t)$ from the primary signal $x(t)$ provides a residual $e(t)$ which controls the process to update the weights $\mathbf{w}$ of the FIR filter. The method proceeds iteratively. The diagonal line across the FIR filter block denotes that the parameters of the block (i.e. $\mathbf{w}$) are updated, but are not part of the signal processed by the block.}
	\label{fig:arlsBlock}
\end{figure}

ARLS proceeds as follows:

\begin{enumerate}
	\item Initialise the tap weights $\mathbf{w} = \mathbf{0}$. Initialise a covariance matrix ${\bf P} = \langle {\bf w} {\bf w}^{\rm T} \rangle = \delta^{-1} {\bf I}$ of rank $M$ for regularisation parameter $\delta$ and an $M \times M$ identity matrix $\mathbf{I}$ where the superscript $\rm T$ symbolises transposition.
	\item For $1 \leq k \leq K$:
	\begin{enumerate}
		\item Estimate the clutter $\hat{c}_k$ from Equation \eqref{clutter_estimate}
		\item Calculate the residual $e_k$ from Equation \eqref{eq:error_estimate}
		\item Calculate the gain vector,
		\begin{eqnarray}
			\mathbf{g}_k = \frac{\mathbf{P} \mathbf{u}_k}{\lambda + \mathbf{u}_k^{\rm T}\mathbf{P} \mathbf{u}_k} \ .
		\end{eqnarray}
	\item Update the tap weights,
			\begin{eqnarray}
		\mathbf{w}_k = \mathbf{w}_{k-1} +  e_k \mathbf{g}_k  \ .
	\end{eqnarray}
	\item Update the covariance matrix,
\begin{eqnarray}
	\mathbf{P}_k =  \mathbf{P}_{k-1}  \lambda^{-1}  - \mathbf{g}_k \mathbf{u}_k^{\rm T}  \lambda^{-1} \mathbf{P}_{k-1} \ .
\end{eqnarray}

	\end{enumerate}

\end{enumerate}
The pseudocode is depicted in Figure \ref{fig:arlsBlock} as a block diagram. At time $t$ the filter receives the reference signal $r(t)$ and produces an estimate of the clutter $\hat{c}(t)$, given the current tap weights $\mathbf{w}$. By comparing $\hat{c}(t)$ with the strain data $x(t)$ a residual $e(t)$ is calculated. The residual is then used in conjunction with $r(t)$ to update the weights and the algorithm continues iteratively. \newline 

ARLS has three free parameters: the order parameter $M$, the forgetting factor $\lambda$, and the regularisation parameter $\delta$. Larger values of $M$ increases the model's complexity. Whilst this generally increases the accuracy of the resulting estimates, it also increases the computational overhead and hence the latency between receiving the input data and updating the weights. In Sec.~\ref{sec:results} we trial a selection of $M$ values for synthetic GW data. Regarding the forgetting factor,  $\lambda=1$ corresponds to infinite memory, whereupon the filter becomes a ``growing window" RLS \citep{10.5555/547203,10.5555/560138}. In this work we use $\lambda=0.9999$ as a default. Different values of $\lambda$ are investigated in Section \ref{sec:roc2}. The regularisation parameter $\delta$ must be chosen judiciously to balance competing priorities  \citep{ljung1999system,1989system}. High $\delta$ leads to rapid convergence of the ARLS algorithm, but with correspondingly high variance in the parameter estimates before convergence. Conversely, lower $\delta$ leads to slower convergence and smaller variance in the parameter estimates. General guidelines for choosing $\delta$ are not readily available. Instead $\delta$ is typically selected empirically for the particular application. In this work we find $\delta = 100$ to be a reasonable choice, given the uncertainty in our initial estimate of the weights. We refer the reader to Chapter 9 of Ref.~\cite{HaykinAdaptiveFT:2002} for a full review of adaptive least squares estimation in the context of linear filtering. 

\section{Validation tests with an HMM} \label{sec:results}
We test the ANC scheme in Section \ref{sec:method} by combining it with a HMM to search for a CW signal with randomly wandering frequency, which overlaps the mains power line. Many other validation tests are possible, of course, and no individual test represents the final word on the efficacy of the scheme. Nonetheless, HMM-based searches are a key element of the data analysis program of the LVK Collaboration, generating a substantial corpus of published results \citep{LIGOMarkov17,LIGOMarkov19,PhysRevD.99.084042,2018PhRvD..97d3013S,LIGOMarkov22,2023PhRvD.107f4062V}. In the context of this paper, validations tests involving a HMM are especially powerful, because they present the ANC scheme with an additional challenge: to detect an underlying CW signal, which is itself ``noisy'' (in the sense that its frequency wanders) rather than monochromatic. That is, the ANC-HMM combination is tasked to not only remove the stochastic 60-Hz interference but also distinguish it from the process noise in the CW signal's frequency. If the ANC succeeds in this complicated task, one may be confident that it also succeeds in the easier task of detecting an underlying CW signal whose frequency is constant.  \newline 

In what follows, we employ the HMM scheme introduced by \citet{Suvorova2016PhRv}, which has been thoroughly tested through multiple LVK searches \cite{Piccinni2022,Riles2023,Wette2023}. We do not cover any details of the HMM scheme in this work and refer the reader to \citet{Suvorova2016PhRv} for further information. A synthetic CW signal $h(t)$, whose frequency executes an unbiased random walk, is injected into synthetic Gaussian noise $n(t)$ typical of the LIGO detectors. Clutter $c(t)$ proportional to a simulated 60-Hz mains power line is superposed, whose properties mimic the real electrical supply (see Section \ref{sec22}). The challenge is to recover $h(t)$ by passing the data $x(t)$ through the ANC filter and feeding the output into the HMM tracker. In Section \ref{sec:creating_data} we describe how we create a synthetic time series $x(t)$. In Section \ref{sec:representative_example} we present a representative example and compare the HMM's ability to detect and track $h(t)$ before and after applying ANC. 

\subsection{Synthetic data} \label{sec:creating_data}
In the tests that follow, we work with representative synthetic data, generated under controlled conditions, rather than directly with LIGO data. We generate synthetic data for the constituents of Equation \eqref{eq:data}, namely $h(t)$, $c(t)$, and $n(t)$, by applying the recipe in Section \ref{sec21} as follows. \newline 

We assume that the source is isolated (i.e. not in a binary) and that the GW signal is quasimonochromatic (i.e. the frequency varies slowly over many wave cycles). We also place the source at a fixed sky position and neglect for simplicity the Doppler frequency modulations induced by the rotation and revolution of the Earth. The latter corrections are included readily in searches with real data through standard continuous wave detection pipelines \cite{Suvorova2016PhRv,2022PhRvD.106j2008A}. We note that the Doppler modulations can in principle influence the time a GW signal spends overlapping with the instrumental line, which could influence the performance of the ANC/HMM scheme. Specifically, for a GW source with intrinsic frequency $f_0$, the frequency evaluated in the detector frame is 
\begin{equation}
	f(t) = f_0 \left(1 + \frac{\vec{v}(t)}{c} \cdot \vec{n}\right) \, ,
\end{equation}
for GW source sky position $\vec{n}$ and detector velocity $\vec{v}$. The detector velocity terms includes both the diurnal and annual motions, and are typically of order $\mathcal{O} \left(10^{-6}\right)$ and $\mathcal{O} \left(10^{-4}\right)$ respectively. These modulations are small and generally narrower than the width of the 60 Hz instrumental line. We are therefore justified in neglecting the Doppler effects for this introductory study. Under the assumptions of source quasimonochromaticity, and neglecting the Doppler modulations, the GW model reduces to 
\begin{equation}
	h(t) = h_0\sin[2\pi \phi_{\rm gw}(t)] \ , \label{eq:ht}
\end{equation}
where $h_0$ is the constant amplitude, $\phi_{\rm gw}(t)$ is a random phase,
\begin{equation}
	\phi_{\rm gw}(t) = \int_{0}^{t} ds \, f_{\rm gw}(s)  \ ,
\end{equation}
and the GW frequency $f_{\rm gw}(t)$ evolves stochastically and piecewise linearly from one time step to the next according to
\begin{eqnarray}
	f_{\rm gw}(t_{n+1}) = f_{\rm gw}(t_n) + \epsilon_n 
\end{eqnarray}
where $\epsilon_n$ is a zero-mean Gaussian random variable with variance $\sigma_f^2$, viz.
\begin{equation}
	\epsilon_n \sim \mathcal{N}(0, \Delta t  \, \sigma_f^2) \ . \label{eq:gwfreqnoise}
\end{equation}
Hence the synthetic GW signal $h(t)$ is completely described by the parameters $h_0$, $\sigma_f$, and $f_{\rm gw}(t=0)$ \newline 

The clutter and reference signals evolve according to Equations \eqref{eq:voltage} and \eqref{eq:vclutter} respectively. For this initial study we take $A_r(t) = a_r$, $A_c(t) = a_c$, and $P(t) = P$ to be constant. Different $P$ values are explored in Section \ref{sec:roc1}. Under these assumptions, Equations \eqref{eq:voltage} and \eqref{eq:vclutter} reduce to 
\begin{align}
 	r(t) = & a_r \cos \left[ 2 \pi f_{\rm ac} t +2 \pi \Delta f_{\rm ac} \cos\left(\frac{2 \pi t}{P}\right) + n_{\Theta} (t_n) \right] \nonumber \\ 
 	&+n_r(t_n) \ , \\ 
	c(t) = & a_c \cos \biggl \{ 2 \pi f_{\rm ac} (t_n - \tau_{\rm delay}) \nonumber \\ 
	&+ 2 \pi \Delta f_{\rm ac} \cos \biggl [ \frac{2 \pi (t_n - \tau_{\rm delay)}}{P} \biggr ] \nonumber \\ 
	&+ n_{\Theta} (t_n - \tau_{\rm delay}) \biggr \} \ .
	\label{eq:clutterequation}
\end{align}
The synthetic reference and clutter data are fully described by the parameters $a_r,a_c,P, f_{\rm ac}, \Delta f_{\rm ac},\sigma^2_{\Theta}$, $\sigma_r^2$, $\tau_{\rm delay}$. \newline 

The 12 free parameters are summarised in Table \ref{tab:parameterdescription1}, along with their chosen injected values. There are three amplitude parameters $h_0, a_r, a_c$, with $h \ll a_r, a_c$. There are four noise parameters $\sigma_n, \sigma_{\Theta}, \sigma_r$ and  $\sigma_f$. In this paper $\sigma_n$ and $\sigma_r$ have fixed, constant values, whilst different values for $\sigma_f$ and $\sigma_{\Theta}$ are investigated in Section \ref{sec:representative_example} and Section \ref{sec:roc1} respectively. There are three additional parameters that specify the modulation of the central frequency: $f_{\rm ac}$, $\Delta f_{\rm ac}$ and $P$. In this paper $f_{\rm ac}$ is fixed at 60 Hz. Different values of $\Delta f_{\rm ac}$ and $P$ are investigated in Section \ref{sec:roc1}. Finally there is the initial GW frequency, $f_{\rm gw}(t=0)$, as well as $\tau_{\rm delay}$.
\begin{table*}
	
	\begin{tabular}{llcc}
		\hline 
		
		 \multirow{2}{*}{Parameter} & \multirow{2}{*}{Physical meaning}  & \multicolumn{2}{c}{Injected value} \\
		\cline{3-4}
			 &  &Section \ref{sec:results} & Section \ref{sec:theroccurves}    \\
		\hline
		$h_0$  &    GW strain amplitude & 0.025& 0.022 \\ 
		$a_r$ & Mains voltage amplitude at PEM & $\mathcal{U}(10-0.01,10+0.01)$& $\mathcal{U}(10-0.01,10+0.01)$ \\
		$a_c$ & Interference voltage amplitude &$\mathcal{U}(1.2-0.001,1.2+0.001)$& $\mathcal{U}(1.2-0.001,1.2+0.001)$ \\
		$\sigma_n$ & Strain channel measurement noise &1& 1 \\
		$\sigma_r$  & PEM voltage measurement noise & $10^{-2}$&$10^{-2}$ \\
		$\sigma_{\Theta}$ & PEM voltage phase noise  & $10^{-1}$& $\{ 10^{-3/2}, 10^{-1}, 10^{-1/2}\}$\\
		$\sigma_f$ &  GW frequency noise & $\{0.1,\sqrt{0.1}\}$ s$^{1/2}$& 0.04 s$^{1/2}$\\
		$f_{\rm gw}(t=0)$ &  Initial GW frequency  & 59.5 Hz& 59.9 Hz\\
		$ f_{\rm ac}$ &  Central interference frequency & 60 Hz& 60 Hz\\ 
		$\Delta f_{\rm ac}$ &  Amplitude of interference frequency modulation  & 1 Hz& $\{0.5, 1.0, 1.5\}$ Hz\\
		$P$ & Period of interference frequency modulation & $10^2$ s& $\{ 10^{1}, 10^2, 10^3\}$ s \\
		$\tau_{\rm delay}$ & Time delay between interference and reference & $\mathcal{U}(0,100 \Delta t)$ s& $\mathcal{U}(0,100 \Delta t)$ s \\
		\hline
	\end{tabular} 
	\caption{Parameters, their physical meaning, and the injected values used to generate synthetic data. The injected values are listed for validating the ANC scheme in Section \ref{sec:results}, and quantifying the performance of the scheme as a function of the properties of the mains power interference in Section  \ref{sec:theroccurves}. $a_r$ and $a_c$ are randomly drawn from separate uniform distributions, denoted by $\mathcal{U}$. The braces notation (e.g. $\{0.01, 0.1\}$) indicates that multiple injected values are trialled. The quoted injected values of the three amplitude parameters and the four noise parameters are scaled such that $\sigma_n^2 = 1$.}
	\label{tab:parameterdescription1}
\end{table*}

\subsection{Representative worked examples} \label{sec:representative_example}
\begin{figure}
	\begin{center}
		\includegraphics[width=\columnwidth]{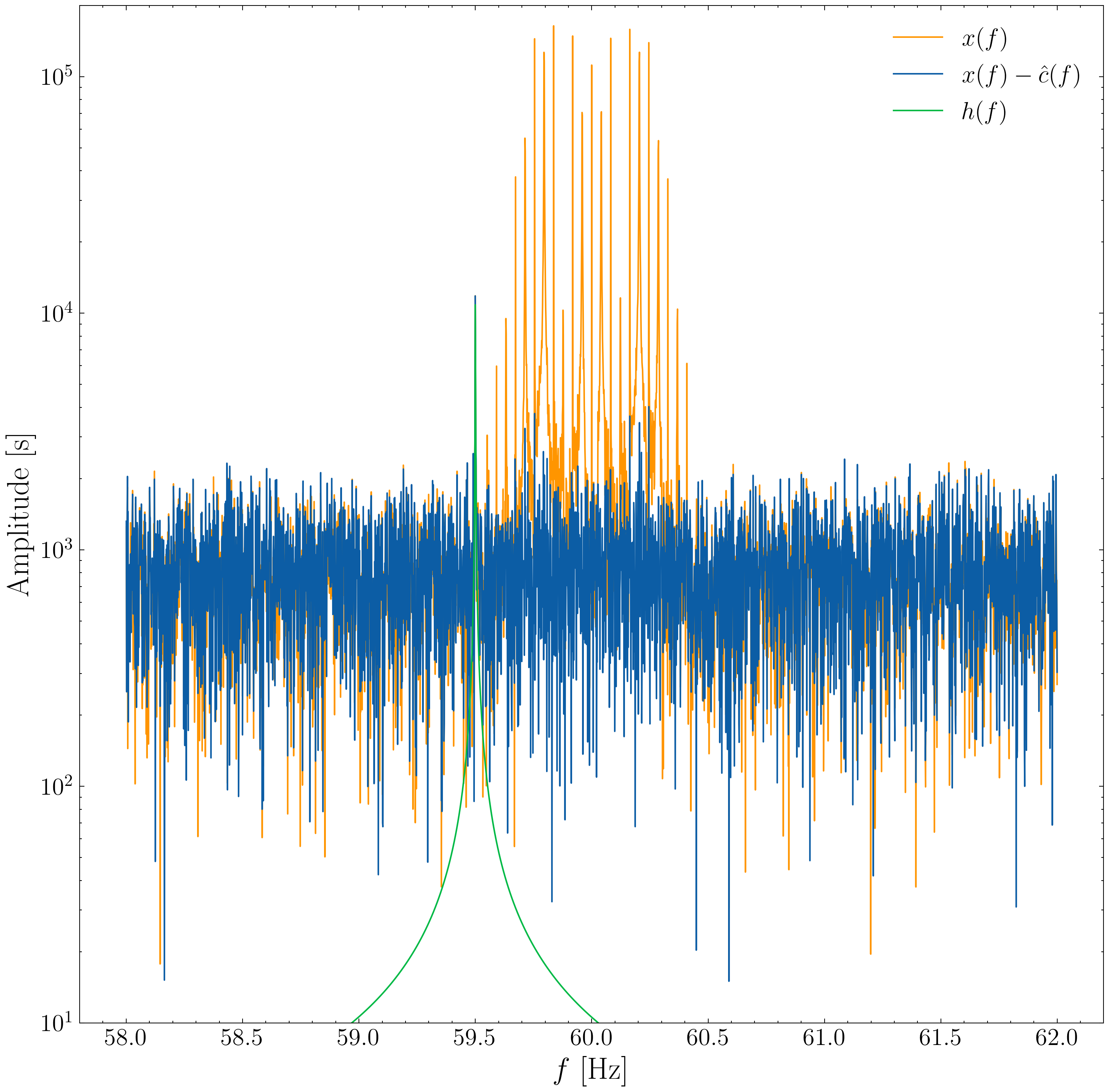}
	\end{center}
	\caption{Spectrum (i.e.\ modulus of the Fourier transform) of the data $x(t)$ (orange curve), GW signal $h(t)$ (green curve) and decluttered output of the ANC filter $x(t) - \hat{c}(t)$ (blue curve) for a system with parameters described in Table \ref{tab:parameterdescription1}. The peak in the blue curve at  59.5 Hz is partially obscured by the green curve. ANC suppresses by $\approx 40\, {\rm dB}$ the orange spikes near $60 \, {\rm Hz}$, corresponding to mains power interference.}
	\label{fig:spectrum}
\end{figure}
\begin{figure*}
	\begin{center}
			\includegraphics[width=\textwidth]{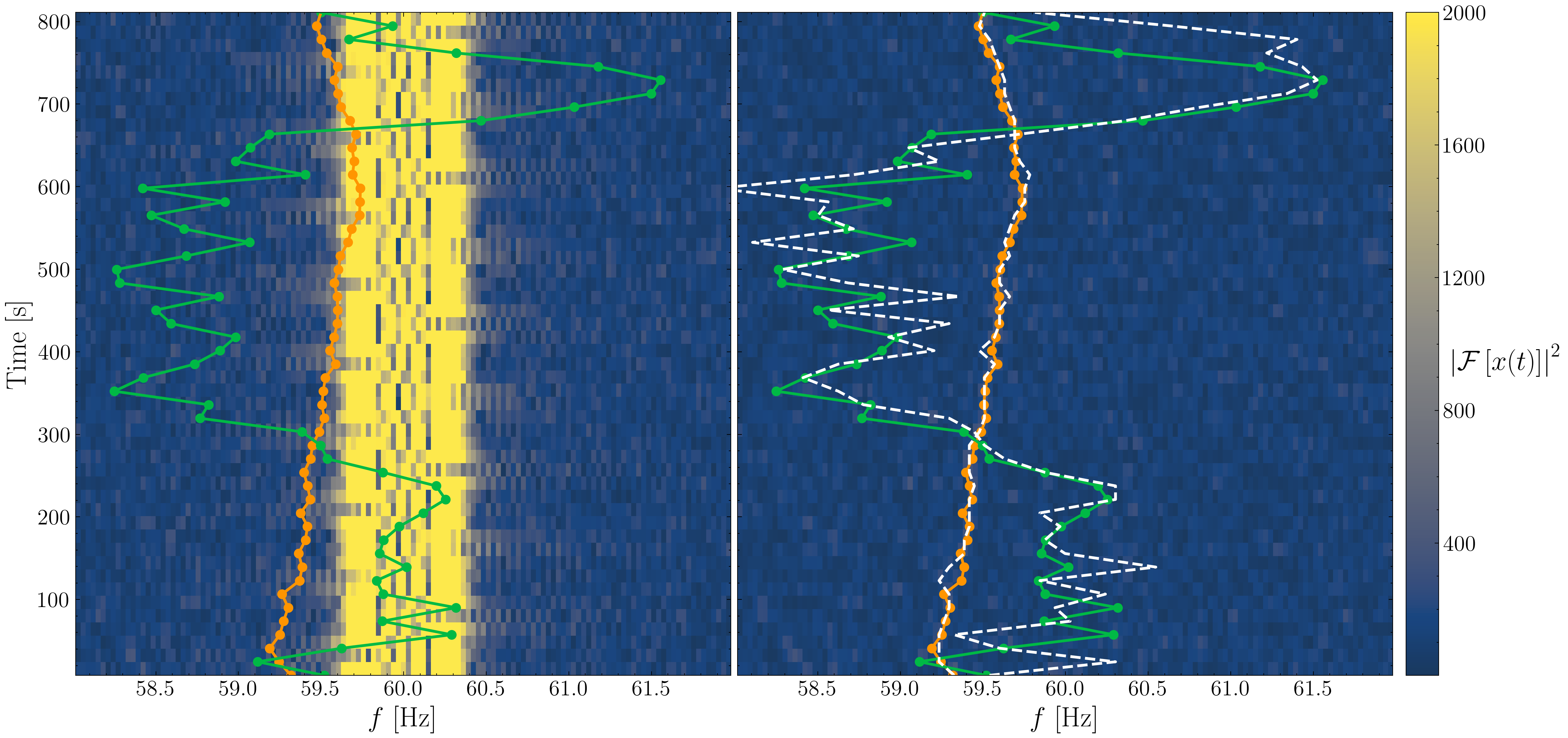}
		\end{center}
	\caption{\label{frequency tracking before and after1}
			HMM tracking of a CW signal with spin wandering in the presence of $60 \, {\rm Hz}$ mains power interference before (left panel) and after (right panel) ANC. Coloured contours (arbitrary units; see colour bar at right) show the spectrogram of the data, i.e.\ the squared modulus of the Fourier transform of $x(t)$. The $60 \, {\rm Hz}$ line is visible as a vertical yellow band in the left panel; it cannot be discerned in the right panel. Injected signals: lower spin wandering, with $\sigma_f (N \Delta t)^{1/2} / f_{\rm ac} = 0.047$ and $ h_0 \sqrt{T_{\rm obs}} / \sigma_n = 0.71$ (solid orange curve); higher spin wandering, with $\sigma_f (N \Delta t)^{1/2} / f_{\rm ac} = 0.15$ and $ h_0 \sqrt{T_{\rm obs}} / \sigma_n = 0.71$  (solid green curve). Recovered signals: dashed white curves. Neither signal can be detected before ANC, so the dashed white curves are visible in the right panel only.}
\end{figure*}
To orient the reader, we start with two representative examples where the frequency wandering is  (i) `low', with $\sigma_f (N \Delta t)^{1/2} / f_{\rm ac} = 0.047$, and (ii) `high', with $\sigma_f (N \Delta t)^{1/2} / f_{\rm ac} = 0.15$. In both cases the effective signal-to-noise ratio equals $ h_0 \sqrt{T_{\rm obs}} / \sigma_n = 0.71$, where $T_{\rm obs}=800 s$ is the total observation time (cf. vertical axis of Figure \ref{frequency tracking before and after1}).  All other parameters are specified in Table \ref{tab:parameterdescription1}. For clarity of exposition, we work with a single PEM in this section and extend to multiple PEMs in Section \ref{sec:roc2}. \newline

We start by checking visually that the ANC filter removes most of the excess power from the 60 Hz interference. In Figure \ref{fig:spectrum} we plot the modulus of the Fourier transforms of the synthetic data $x(t)$, the underlying GW signal $h(t)$, and the decluttered output of the ANC filter, $e(t) = x(t) - \hat{c}(t)$, for Fourier frequencies from $58 \, {\rm Hz}$ to $62 \, {\rm Hz}$ for case (i). Before filtering, the spectrum of $x(t)$ has multiple peaks near $f = f_{\rm ac} = 60 \, {\rm Hz}$, visible as the orange spikes in Figure \ref{fig:spectrum}. After filtering, the spectrum of $e(t)$ is flatter, as indicated by the blue curve in Figure \ref{fig:spectrum}, with the exception of a clear feature at $59.5$ Hz coincident with the GW injection at $f_{\rm gw}(t=0)$ (green curve). In rough terms, the ANC scheme achieves $20 \, {\rm dB}$ of suppression in case (i). Similar results (not plotted for brevity) are obtained for case (ii). \newline 
 
Having established the ability of the ANC scheme to filter out $c(t)$ by calibrating against $r(t)$, we pass the ANC output to the HMM tracker and evaluate the tracker's performance. The results are shown in  Figure \ref{frequency tracking before and after1} for both low and high frequency wandering, for a single realisation of the noise. The figure shows the spectrogram of $x(t)$ as a heat map before (left panel) and after (right panel) ANC filtering. It is clear that ANC suppresses the mains power interference, which is visible as a vertical yellow band in the left panel and is almost absent from the right panel. The spin wandering of the injected GW source (green solid curve for higher $\sigma_f$, orange solid curve for lower $\sigma_f$) and the HMM estimate (dashed white curves for both higher and lower $\sigma_f$) are superposed onto the spectrogram. The approximate overlap between the solid coloured and dashed white curves confirms that the ANC scheme and HMM tracker detect both injected signals successfully. For lower $\sigma_f$, the GW spin frequency wanders close to, but below, the 60 Hz interference line. For higher $\sigma_f$, the GW signal crosses the interference line, presenting a more difficult challenge, which is surmounted successfully nevertheless. The time-averaged root-mean-square error in the frequency estimate is $0.038$ Hz for low $\sigma_f $ and $0.47$ Hz for high $\sigma_f$. In contrast, neither the lower-$\sigma_f$ nor the higher-$\sigma_f$ injections are detected by the HMM tracker before ANC in the left panel (which is why there are no dashed white curves in the left panel). We refer the reader to \cite{Suvorova2016PhRv} and \cite{2017PhRvD..96j2006S} for a discussion of the accuracy of HMM tracking generally, and the magnitude of the time-averaged root-mean-square error in the frequency estimates without ANC ($\lesssim 1/(2\Delta t) = 0.03$ Hz here).

\section{ROC curves}\label{sec:theroccurves}

%In Section \ref{sec:roc1} we quantify the performance of the ANC filter as a function of the properties of the mains power interference (e.g.\ $\Delta f_{\rm ac}$). In Section \ref{sec:roc2} we quantify the performance as a function of the ANC filter's settings, e.g. its order $M$ (i.e. the number of taps) and the number of PEM inputs.

In this section, we quantify the performance of the HMM tracker, when it analyzes filtered data supplied by the ANC scheme. We do so systematically by computing receiver operating characteristic (ROC) curves as a function of key parameters of the mains power interference (Section \ref{sec:roc1}) and ANC filter (Section \ref{sec:roc2}).

\subsection{Mains power interference parameters} \label{sec:roc1}

\begin{figure}
	\begin{center}
		\includegraphics[width=\columnwidth]{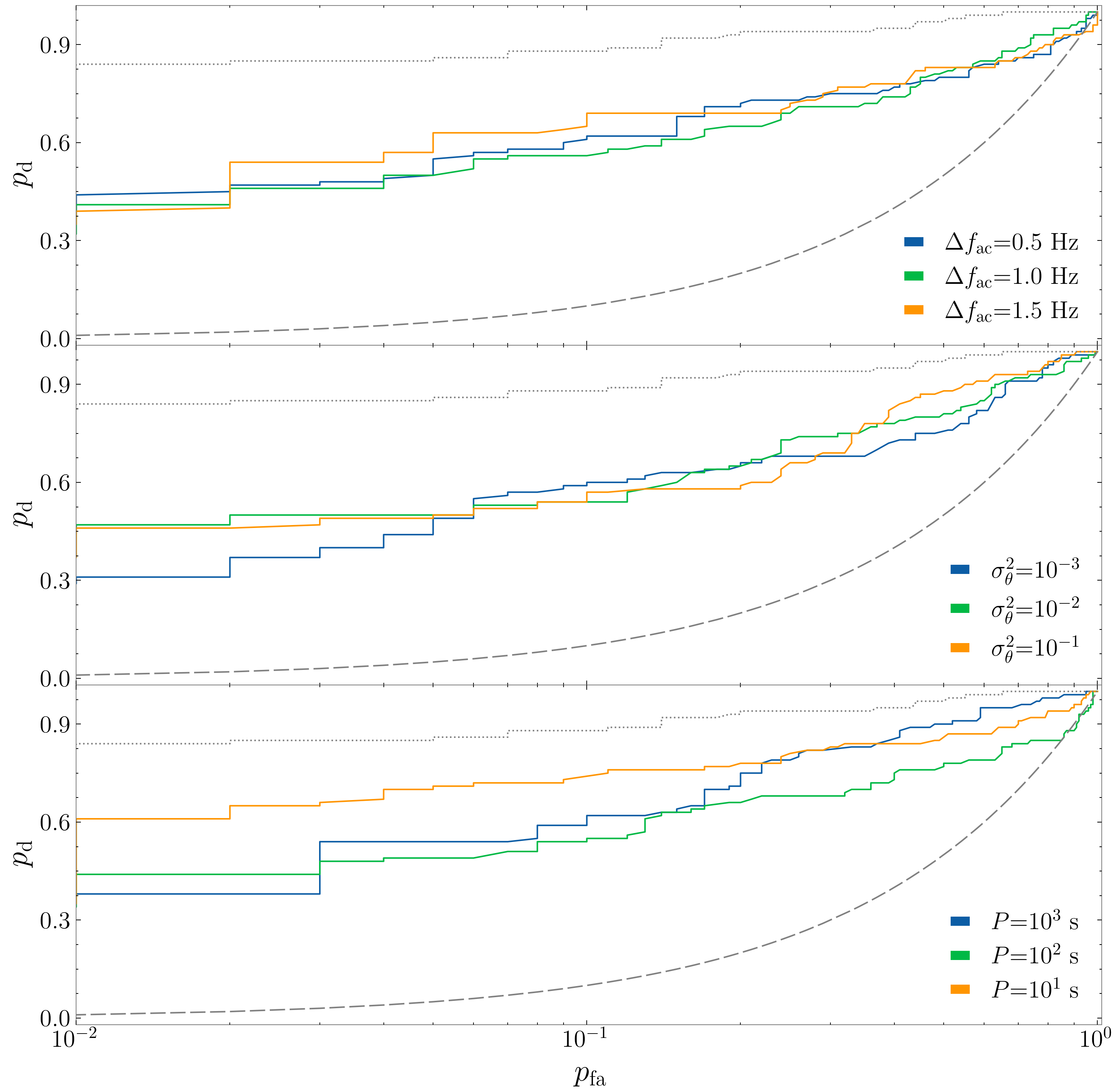}
	\end{center}
	\caption{Detection probability $p_{\rm d}$ as a function of false alarm probability $p_{\rm fa}$ (i.e. ROC curve) for HMM tracking of a CW signal ($ h_0 \sqrt{T_{\rm obs}} / \sigma_n = 0.62$) in conjunction with the ANC scheme, for different mains power interference parameters. The top, middle and bottom panels display different values of $\Delta f_{\rm ac}$, $\sigma_{\Theta}^2$, and $P$ respectively. The grey dashed curve in all panels corresponds to the performance of a random classifier. The grey dotted curve in all panels is a performance upper limit, corresponding to the case where there is zero mains power interference. In all panels the curves overlap approximately, illustrating that the performance of the HMM and ANC scheme is insensitive to the mains power parameters. At $p_{\rm fa} = 0.05$, the mean $p_{\rm d}$ across the three curves is 0.52 (top panel), 0.48 (middle panel) and 0.58 (bottom panel).}
	\label{fig:roc1}
\end{figure}
In this section we test how the ANC scheme and HMM tracker perform together as a function of the mains power parameters $\Delta f_{\rm ac}$, $\sigma_\Theta^2$ and $P$. To this end, we calculate the probability of detecting the CW signal, $p_{\rm d}$, as a function of the false alarm probability, $p_{\rm fa}$, i.e. the ROC curve $p_{\rm d} (p_{\rm fa})$. We consider three numerical experiments. In the first experiment we vary $\Delta f_{\rm ac} =  \{0.5, 1.0, 1.5\}$ Hz whilst holding constant $\sigma_{\Theta}^2 = 10^{-2}$ and $P = 100 $ s. In the second experiment we vary $\sigma_{\Theta}^2 =  \{ 10^{-3}, 10^{-2}, 10^{-1}\}$ whilst holding constant $\Delta f_{\rm ac} = 1.0$ Hz and $P = 100 $ s. In the third experiment we vary $P=  \{ 10, 10^2, 10^3\}$ s whilst holding constant $\Delta f_{\rm ac} = 1.0$ Hz and $\sigma_{\Theta}^2 = 10^{-2}$. For all experiments we fix $ h_0 \sqrt{T_{\rm obs}} / \sigma_n = 0.62$, $f_{\rm gw}(t=0) = 59.9$ Hz, and $N \Delta t = 800$ s. Despite expecting $\Delta f_{\rm ac} < 0.5$ in practice, as discussed in Section \ref{sec21}, we trial larger values of $\Delta f_{\rm ac}$ in order to set a more difficult test for the ANC scheme and HMM tracker. All injected parameters are specified in Table \ref{tab:parameterdescription1}. \newline

Figure \ref{fig:roc1} displays the ROC curves resulting from the three numerical experiments above in the top, middle and bottom panels respectively. For example in the top panel the results of the first experiment are plotted, where we set  $\Delta f_{\rm ac} = \{0.5, 1.0, 1.5\}$ Hz (blue, green and orange curves respectively). Also plotted for comparison is the performance of a random classifier (grey dashed curve). The performance across different $\Delta f_{\rm ac}$, $\sigma_\Theta$, and $P$ values is broadly comparable, with the ROC curves overlapping approximately and outperforming a random classifier (i.e. each of the ROC curves lies above the grey dashed line). The performance is not as good as in the zero interference case (i.e. each of the ROC curves lies below the grey dotted line) because ANC is imperfect of course. The performance of the ANC filter can be improved by tuning the relevant parameters, which is discussed in Section \ref{sec:roc2}. \newline

We take $p_{d}(p_{\rm fa} =0.05)$ as a scalar metric to quantify the typical performance of the HMM-ANC combination in practical applications. In the top panel we obtain $p_{\rm d}(0.05$) = $ \{0.50, 0.50, 0.57 \}$ for $\Delta f_{\rm ac} =  \{0.5, 1.0, 1.5\}$ Hz respectively. In the middle panel we obtain $p_{\rm d}(0.05$) = $ \{0.44,0.50,0.49 \}$ for $\sigma_{\Theta}^2 =  \{ 10^{-3}, 10^{-2}, 10^{-1}\}$ respectively. In the bottom panel we obtain $p_{\rm d}(0.05) = (0.7,0.49,0.54)$ for $P=  \{ 10, 10^2, 10^3\}$ s respectively.

\subsection{ANC filter parameters} \label{sec:roc2}
\begin{figure}
	\begin{center}
		\includegraphics[width=\columnwidth]{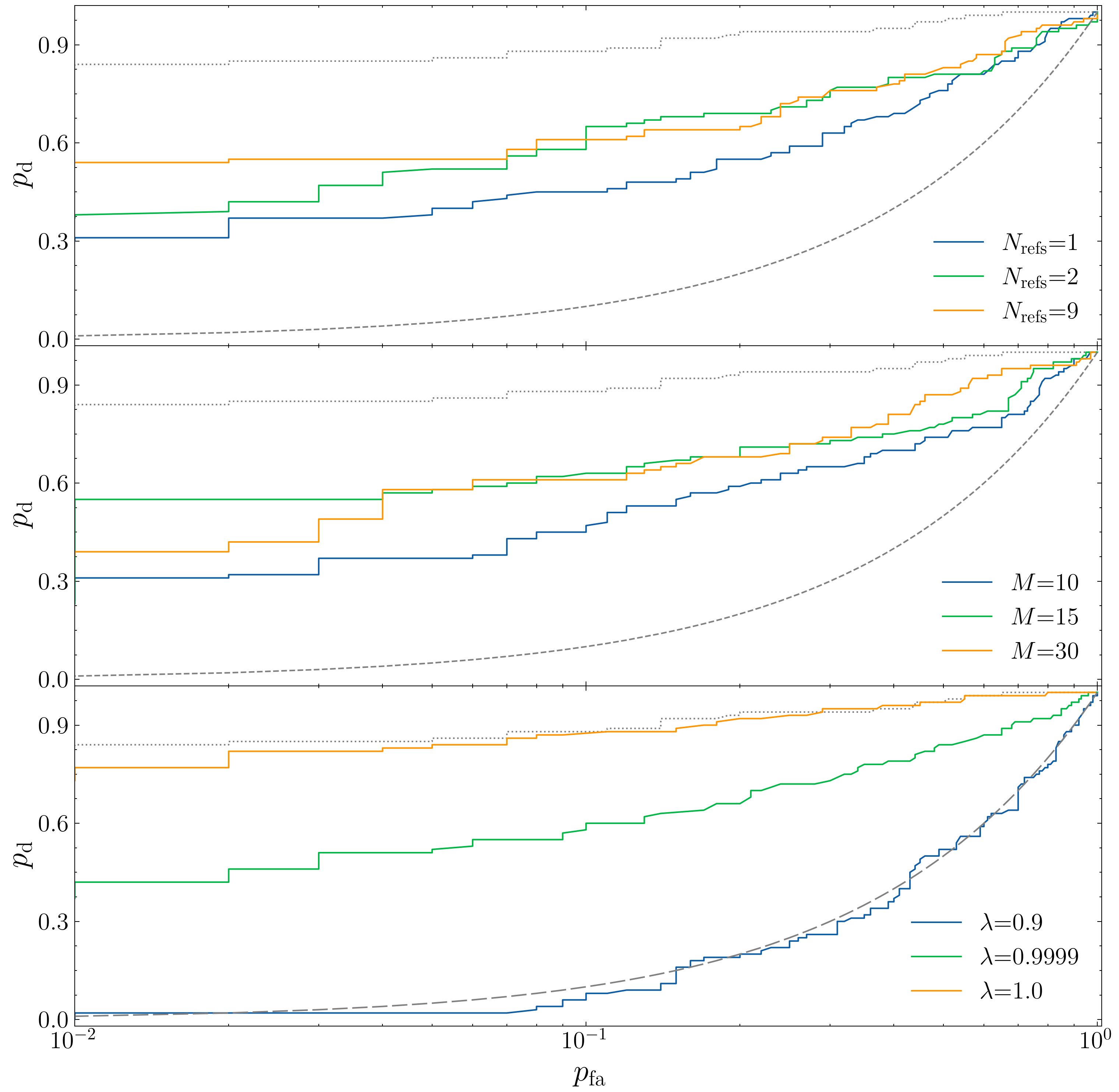}
	\end{center}
	\caption{Same as Figure \ref{fig:roc1} but for different ANC filter parameters. The top, middle and bottom panels display the ROC curves for different numbers of PEM reference channels $N_{\rm ref}$, tap weights $M$, and forgetting factors $\lambda$ respectively. All panels additionally include the ROC curve for a random classifier (i.e. the lower performance limit, grey dashed curve) and the zero interference ROC curve (i.e. the upper performance limit, grey dotted curve).}
	\label{fig:roc2}
\end{figure}
We verify in Section \ref{sec:roc1} that the ANC scheme and HMM tracker are insensitive to the mains power interference parameters. In this section we investigate the performance as a function of the control parameters of the ANC filter itself. Specifically, we investigate three important questions:
\begin{enumerate}
	\item How does ANC benefit from multiple independent references?
	\item What order ANC filter ($M$) is required to achieve good interference cancellation? 
	\item How does ANC performance depend on $\lambda$?
\end{enumerate}

Figure \ref{fig:roc2} displays the ROC curves for different values of the number of PEM references ($N_{\rm ref}$, top panel), $M$ (middle panel) and $\lambda$ (bottom panel). All mains power interference parameters are as specified in Table \ref{tab:parameterdescription1}. Also plotted for comparison in every panel is the ROC curve of a random classifier (grey dashed curve) and the ROC curve for $c(t) = 0$ (grey dotted curve). The zero interference case is included as a performance upper bound. \newline 

The validation tests in Sections \ref{sec:representative_example} and \ref{sec:roc1} assume a single PEM reference voltage. In practice, as discussed in Section \ref{sec23}, there are multiple PEM channels measuring power line interference for LIGO (c.f. Figure \ref{coherenceplot_1}). Specifically, for O3 there are nine PEM channels which directly measure the three-phase voltages at each of the LIGO-Livingston and LIGO-Hanford sites. Multiple PEM channels provide additional independent references, which should improve the performance of the ANC filter \footnote{ANC is commonly used with multiple reference signals in other electrical engineering applications such as noise cancelling headphones \citep{10.1121/1.5109394}, communication intelligibility \citep{KUO1996669,doi:10.1177/1084713812456906} and cardiac monitoring \citep{7755741}.}. Indeed an improvement of this kind is visible in the top panel of Figure \ref{fig:roc2} which plots the ROC curves for $N_{\rm ref} = \{ 1,2,9 \}$ (blue, green and orange curves respectively). The detection probability is found to be $p_{\rm pd} (0.05) = \{ 0.38,0.52,0.55 \}$ for the respective $N_{\rm ref}$ values. For comparison, in the zero interference case we obtain $p_{\rm pd}(0.05) = 0.85$. It is evident that $N_{\rm ref} > 1$ leads to an improvement over $N_{\rm ref} = 1$. However, the improvement of $N_{\rm ref} =9$ over $N_{\rm ref} =2$ is more modest for $p_{\rm fa} \gtrsim 0.04$. This suggests that the inclusion of additional reference PEM channels reaches a point of diminishing returns. \newline

With respect to the second question concerning the order of the filter, more taps should improve performance, as the complexity of the model increases. Indeed, this trend is observed in the middle panel of Figure \ref{fig:roc2}  which plots ROC curves for $M = \{10,15,30\}$ (blue, green, and orange curves respectively). The detection probability is found to be $p_{\rm pd} (0.05) = \{0.37, 0.57, 0.58\}$ for each respective $M$ value. Analogous to the top panel, increasing $M$ generally increases the detection probability, but the performance saturates; the ROC curves for $M=15$ and $M=30$ are comparable for $p_{\rm fa} \gtrsim 0.04$. More taps increases the computational costs and latency (i.e. the delay in the signal processing pipeline due to the filter). In this work we explicitly specify the number of taps, while noting that adaptive tap length methods are also available \citep[e.g.][]{1326385,KAR2017422,KAR2020107043}. \newline

With respect to the third question concerning the forgetting factor $\lambda$, we find that $p_{\rm d}$ is a strong function of $\lambda$ at fixed $p_{\rm fa}$. In the bottom panel of Figure \ref{fig:roc2}, we plot ROC curves for $\lambda = \{0.90,0.9999,1.0\}$ (blue, green, orange curves respectively), which have $p_{\rm pd} (0.05) = \{0.02, 0.51, 0.83\}$ respectively. A clear hierarchy is evident whereby $p_{\rm d}$ increases with $\lambda$. For the synthetic data in this paper $\lambda=1$ is advantageous for detection since the interference is quasi-stationary (c.f. Equation \ref{eq:clutterequation}). Consequently, discounting older data with $\lambda < 1$ inhibits the ability of the ANC scheme to filter out the interference. We defer investigation of non-stationary interference (c.f. Equation \ref{eq:vclutter}) to future work. When analysing real astrophysical data $\lambda$ should be selected optimally. Algorithms exist to achieve this, as in Ref. \citep{6349749}. For $\lambda = 1$, ARLS has infinite memory and so exhibits good stability (i.e. insensitivity to numerical perturbations) and low misadjustment (i.e. $\mathbf{w}$ is not erroneously updated away from its optimal value $\mathbf{w}_{\rm opt}$). Conversely, $\lambda < 1$ improves the tracking at the expense of reduced stability and increased misadjustment \citep{Ciochina5206117}. In this paper we run ARLS using a specific value for $\lambda$, but note that adaptive $\lambda$ algorithms are also available \citep{app12042077,1468506,4639569}. \newline

\section{Real LIGO noise}\label{sec:real_data}
The success of the ANC filter when applied to synthetic data motivates the extension of the validation tests to real data. In this section we repeat the analysis of Section \ref{sec:results} but work directly with LIGO data. In Section \ref{sec:real_data_generation} we describe the data used for the tests. In Section \ref{sec:real_data_example} we present a representative example, analogous to Section \ref{sec:representative_example}, and compare the HMM's ability to detect and track a synthetic GW signal injected into real data, before and after filtering the data using ANC.

\subsection{Data}\label{sec:real_data_generation}
In order to test the scheme, we require real data from the strain channel, $x'(t)$, real data from the reference channel, $r'(t)$, and an injected GW signal, $h(t)$. We use the primed notation, $x'(t)$ and $r'(t)$, to emphasise that these timeseries are acquired from the LIGO instrument itself rather than from e.g. solving Equations \eqref{eq:data} or \eqref{eq:voltage}. \newline

 We acquire $x'(t)$ from the LIGO-Livingston strain channel \texttt{L1:DCS-CALIB\_STRAIN\_C01\_AR}. We acquire $r'(t)$ from the nine independent PEM channels that directly measure the mains voltage (see Section \ref{sec23} and Figure \ref{coherenceplot_1}). For both $x'(t)$ and $r'(t)$ we use data from the first part of the third LIGO observing run, O3a \cite{O3data2023}, obtained via the LIGO Data Grid. The strain data are sampled at 16384 Hz and downsampled to 1024 Hz to match the sampling rate of the PEM channels. As no continuous wave signal has yet been detected in the LIGO data, we retain a synthetic injected GW signal for $h(t)$, via Equations \eqref{eq:ht}--\eqref{eq:gwfreqnoise}. The total strain data which is ingested by the ANC scheme is then $x'(t) + h(t)$.

\subsection{Representative worked example}\label{sec:real_data_example}
\begin{figure}
	\begin{center}
		\includegraphics[width=\columnwidth]{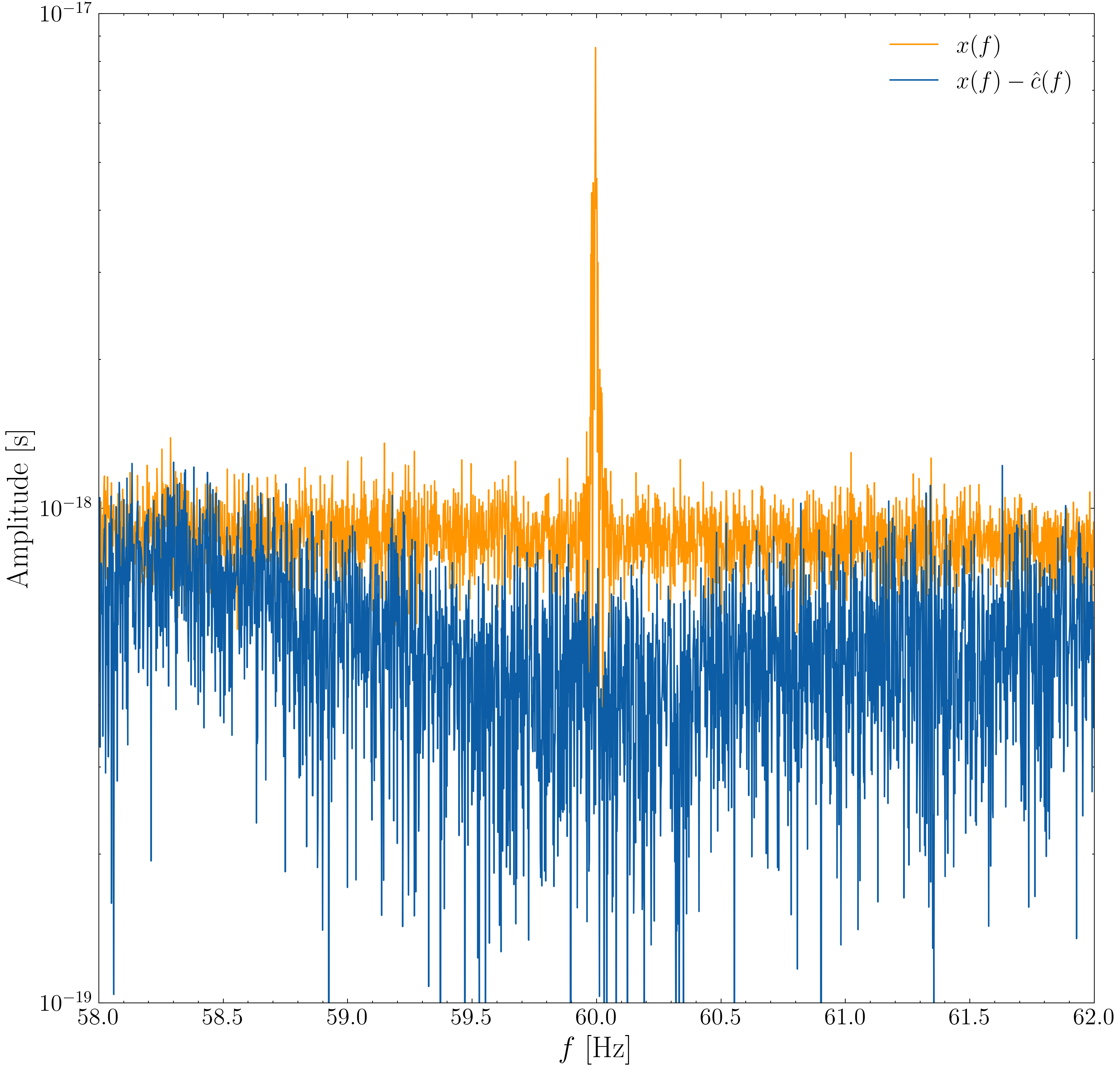}
	\end{center}
	\caption{Spectrum (i.e.\ modulus of the Fourier transform) of the data $x'(t)$ (orange curve), and decluttered output of the ANC filter $x'(t) - \hat{c}(t)$ (blue curve) for LIGO data obtained via the procedure outlined in Section \ref{sec:real_data_generation}. ANC suppresses by $\approx 20\, {\rm dB}$ the orange spike near $60 \, {\rm Hz}$, corresponding to mains power interference. Fig. \ref{fig:spectrum} is the same as Fig. \ref{fig:spectrum_real} but for synthetic noise generated via the procedure in Section \ref{sec:real_data_generation}.}
	\label{fig:spectrum_real}
\end{figure}
\begin{figure*}
	\begin{center}
		\includegraphics[width=\textwidth]{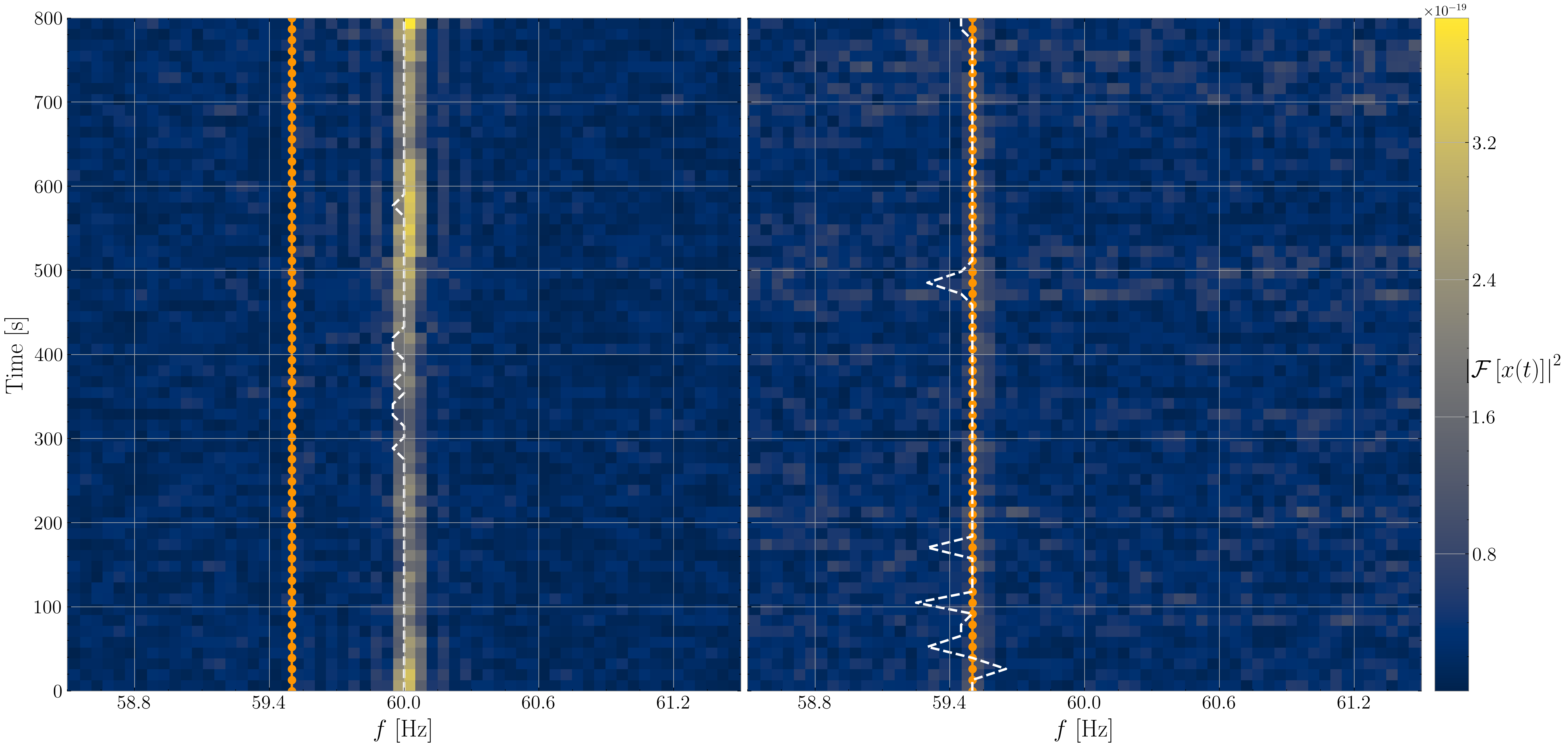}
	\end{center}
	\caption{\label{frequency tracking before and after_real_Date} Same as Figure \ref{frequency tracking before and after1}, but for a GW signal (orange curve) injected into real LIGO noise. The $60 \, {\rm Hz}$ line is visible as a vertical yellow band in the left panel; it cannot be discerned in the right panel. The signal recovered by the HMM is given by the dashed white curves. Before ANC, the HMM erroneously tracks the 60 Hz interference line. After ANC, the HMM successfully tracks the underlying GW signal.}
\end{figure*}

To start, we confirm visually that the ANC filter applied to real data removes much of the excess power from the 60 Hz interference. In Figure \ref{fig:spectrum_real} we plot the modulus of the Fourier transforms of the  data $x'(t)$, and the decluttered output of the ANC filter, $e(t) = x'(t) - \hat{c}(t)$, for Fourier frequencies from $58 \, {\rm Hz}$ to $62 \, {\rm Hz}$, analogously to Figure \ref{fig:spectrum}. Before filtering, the spectrum of $x(t)$ exhibits a clear peak near $f = f_{\rm ac} = 60 \, {\rm Hz}$, visible as the orange spike in Figure \ref{fig:spectrum_real}. After filtering, the spectrum of $e(t)$ is flatter, as indicated by the blue curve. In rough terms, the ANC scheme achieves $20 \, {\rm dB}$ of suppression. We note that the ANC marginally over-subtracts the noise. This is likely due to the non-linear relationship between the strain and the PEM channels, cf. the coherence discussion in Section \ref{sec23}. Additionally, the free parameters of the ANC filter (cf. Section \ref{sec:roc2}) could be better tuned. \newline

We now evaluate the performance of the combined ANC-HMM scheme. We inject a GW signal $h(t)$ with a constant frequency $f_{\rm gw}(t) = f_{\rm gw} (t=0) = 59.5 $ Hz. The results are shown in Figure \ref{frequency tracking before and after_real_Date}. The figure is analogous to Figure \ref{frequency tracking before and after1}; 
it show a spectrogram of $x'(t)$ as a heat map before (left panel) and after (right panel) ANC filtering. ANC suppresses the mains power interference, which is visible as a vertical yellow band in the left panel and is absent from the right panel. The injected $h(t)$ (green solid curve) and the HMM estimate of the frequency track (dashed white curve) are superposed onto the spectrogram. Before applying ANC, the 60 Hz line corrupts the performance of the HMM, which erroneously tracks the noise line. After applying ANC the 60 Hz line is removed and the HMM successfully tracks the underlying GW signal (solid coloured and dashed white curves overlap in right-side panel). The time-averaged root-mean-square error in the frequency estimate is $\approx 0.06$ Hz, comparable to the error for synthetic data in the low-$\sigma_f$ regime (see Section \ref{sec:representative_example}).

\section{Conclusions}\label{sec:conclusions}
In this paper we demonstrate a new line subtraction method based on ANC for use in continuous gravitational wave searches. We use an ARLS algorithm in conjunction with an independent, known PEM reference signal to suppress the interference from a long-lived narrow spectral feature. We then search for the continuous wave signal, whose frequency wanders stochastically in general, using a standard HMM search pipeline. We test the scheme on synthetic data modelling the 60 Hz interference from the North American mains power supply. \newline 

We find that ANC suppresses the 60-Hz interference by approximately 40 dB, allowing the HMM to track continuous-wave signals that overlap with the 60-Hz spectral feature. We find the method to be insensitive to different mains power interference parameters, with $p_{\rm d}(0.05) \sim 0.5$ across a range of values of $\Delta f_{\rm ac}$, $\sigma_\Theta$, and $P$. The method does depend on the filter parameters $N_{\rm ref}$, $M$ and $\lambda$. The detection probability $p_{\rm d}(0.05)$ generally increases with these parameters, albeit at a decreasing rate. For the maximum values of the filter parameters trialled, we obtain $p_{\rm d}(0.05) = 0.55, 0.58,0.55$ for $N_{\rm ref} = 9$, $M=30$ and $\lambda=1$ respectively. ROC curves $p_{\rm d}(p_{\rm fa})$ for representative values of $\Delta f_{\rm ac}$, $\sigma_\Theta$, $P$, $N_{\rm ref}$, $M$, and $\lambda$ are presented in Figures \ref{fig:roc1} and \ref{fig:roc2}. \newline 

The method is tested further on real LIGO noise. Before applying ANC, the HMM erroneously tracks the 60-Hz interference line. Applying ANC suppresses the 60-Hz interference by approximately 20 dB and enables the HMM to successfully track the underlying GW signal, with a time-averaged root-mean-square error in the frequency estimate of $\approx 0.06$ Hz (Figures \ref{fig:spectrum_real}  and \ref{frequency tracking before and after_real_Date}). \newline

Ref. \cite{CovasEtAl:2018} catalogues long-lived, narrowband interference in the LIGO interferometers from various sources, including electrical subsystems, mechanical subsystems and calibration processes. Some of these instrumental lines are ``cleaner'' spectrally than mains power interference, while others are not. In the future, it will be interesting to test whether the success achieved by the ANC scheme in this paper extends to other instrumental lines. Such tests involve investment in human and computational resources but, if successful, will increase the proportion of the LIGO band accessible to continuous wave searches at frequencies of astrophysical interest, especially at $\lesssim 0.1 \, {\rm kHz}$.

\bibliographystyle{apsrev4-1} % Tell bibtex which bibliography style to use

\bibliography{bibLinesPaper}
%%%%%%%%%%%%%%%%%%%%%%%%%%%%%%%%%%%%%%%%%%%%%%%%%%

\begin{acknowledgements}
%GWOSC
This material is based upon work supported by NSF's LIGO Laboratory which is a major facility fully funded by the National Science Foundation. This research has made use of data, software and/or web tools obtained from the Gravitational Wave Open Science Center (https://www.gw-openscience.org), a service of LIGO Laboratory, the LIGO Scientific Collaboration and the Virgo Collaboration. LIGO is funded by the U.S. National Science Foundation. Virgo is funded by the French Centre National de Recherche Scientifique (CNRS), the Italian Istituto Nazionale della Fisica Nucleare (INFN) and the Dutch Nikhef, with contributions by Polish and Hungarian institutes. This research also
made use of LIGO Data Grid clusters at the California Institute of Technology. This paper has been
assigned LIGO DCC number P2300372. This research was supported by
the Australian Research Council (ARC) Centre of Excellence for Gravitational Wave Discovery (OzGrav), grant number CE170100004, and ARC Discovery Project DP170103625. HM acknowledges support from the UK Space Agency (grant
nos. ST/Y004922/1; ST/V002813/1; ST/X002071/1). 
\end{acknowledgements}

\end{document}